\documentclass[prd,showpacs,eqsecnum]{revtex4}

\usepackage{amsfonts,amsmath,amssymb,bm,graphicx,subfigure,float}

\begin{document}


\title {Electric charge and magnetic moment of massive neutrino}

\author{Maxim Dvornikov}
\altaffiliation[Also at ]{Department of Theoretical Physics,
Moscow State University, 119992 Moscow, Russia}
\email{maxim_dvornikov@aport.ru}
\author{Alexander Studenikin}
\affiliation{Department of Theoretical Physics,
Moscow State University, 119992 Moscow, Russia}
\email{studenik@srd.sinp.msu.ru}

\date{\today}

\begin{abstract}
We consider the massive Dirac neutrino electric charge and
magnetic moment within the context of the standard model supplied
with $\mathrm{SU}(2)$-singlet right-handed neutrino in arbitrary
$R_{\xi}$ gauge. Using the dimen\-sio\-nal-regularization scheme
we start with the calculations of the one-loop contributions to
the neutrino electromagnetic vertex function exactly accounting
for the neutrino mass. We examine the decomposition of the massive
neutrino electromagnetic vertex function. It is found by means of
direct calculations that the massive neutrino vertex function
contains only the four form factors. Then we derive the closed
integral expressions for different contributions to the neutrino
electric form factor, electric charge, and magnetic moment.
For several one-loop contributions to the neutrino charge and
magnetic moment, that were calculated previously with mistakes by
the other authors, we find the correct results. We show that the
electric charge for the massive neutrino is a gauge independent
and vanishing parameter in the first two orders of the expansion
over the neutrino mass parameter $b=(m_{\nu}/M_{W})^2$. From the
obtained closed two-integral expression for a massive neutrino
electric form factor it is also possible to derive the neutrino
charge radius. In the particular choice of the 't~Hooft-Feynman
gauge we also demonstrate that the neutrino charge is zero in all
orders of expansion over $b$, i.e. for arbitrary mass of neutrino.
For each of the diagrams contributing to the neutrino magnetic
moment, we obtain the expressions accounting for the leading
(zeroth) and next-to-leading (first) order in $b$, where the gauge
dependence is shown explicitly. Each of the contributions is
finite and the sum of all contributions turns out to be
gauge-independent. Our calculations also enable us to obtain the
neutrino magnetic moment in theoretical models that differ from
each other by the values of particles' masses, including the case
of a very heavy neutrino. The general expression for the massive
neutrino magnetic form factor is presented.
\end{abstract}


\maketitle

\section{Introduction}

The recent experimental studies of the astrophysical and
terrestrial neutrino fluxes provide the convincing evidences for
the non-vanishing neutrino mass and neutrino mixing \cite{Bil02}.
These properties of neutrino are attributes of the physics beyond
a scope of the standard model. An important information on the
structure of the future model of interaction can be obtained in
the investigation of radiative corrections to the properties of
neutrino that in principle can be also verified in experiments. A
critical test of a theoretical model is provided by the direct
calculation of such characteristics of neutrino as its electric
charge and magnetic moment. In this respect it is interesting to
examine the gauge and neutrino mass dependence of these
quantities.

The electric charge and magnetic moment are the most important
static electromagnetic properties of a particle. Their values are
determined by corresponding form factors if the external photon is
on a mass shell. In spite of the fact that the electromagnetic
form factors are not measurable properties of a particle at
nonzero momentum transfer, there are processes where the off-shell
external photons are important. The example is the radiative
corrections to the fermion-fermion scattering. The fermion
electromagnetic vertex function and, in particular, its
representation in terms of form factors in the case of off-shell
external photon has been considered in
Refs.~\cite{Kim76,BegMarRud78} within various gauge theories.

To the best of our knowledge, however, there is no direct one-loop
calculations of the neutrino electromagnetic vertex as well as of
the neutrino charge and magnetic moment which are performed within
the context of the standard model in the renormalizable $R_{\xi}$
gauge and which explicitly take into account the neutrino mass. It
is worth to be noted here that the massive Majorana neutrino
cannot have neither magnetic nor electric dipole moment. Due to
the lepton flavour non-conservation that has been confirmed in the
neutrino experiments, a neutrino could have flavour-off-diagonal
transition magnetic moment, which is also allowable for Majorana
neutrino. The differences between the Dirac and Majorana
electromagnetic properties are explained in detail in
\cite{Kay82}.

The neutrino vertex function in the limit of small neutrino mass
was considered in Ref.~\cite{LeeShr77}. There are several works
where the neutrino charge is calculated within the standard model
using the unitary, linear $R_{\xi}$, and 't~Hooft-Feynman gauges
\cite{BarGasLau72,MarSir80,Sak81,LucRosZep84}. The one-loop
contributions to the neutrino magnetic moment in the standard
model have been also considered previously
\cite{LeeShr77,FujShr77,Shr82,EgoLobStu99}. The vanishing of the
massless neutrino charge emerges from the unbroken electromagnetic
gauge invariance. The corresponding Ward identity has been derived
in Refs.~\cite{DenWeiDit95,CabBerVidZep00} using the background
field method. In the recent studies \cite{CabBerVidZep00} a
computation of one-loop electroweak diagrams, which contribute to
the neutrino charge and magnetic moment in the background field
method and in the linear $R_{\xi}$ gauge, is presented. However,
all the previous calculations of the neutrino charge have been
performed under assumption of a vanishing neutrino mass. With
respect to the neutrino magnetic moment, an according treatment of
this quantity was also performed to the leading order in the
neutrino mass that is valid for the case of neutrino being much
lighter than the corresponding charged lepton, $m_{\nu_{\ell}}\ll
m_{\ell}$. In addition, the results presented in
Ref.~\cite{CabBerVidZep00} for the gauge-fixing parameter
dependence of several one-loop contributions to the neutrino
charge and magnetic moment are incorrect.

In this paper we consider the massive Dirac neutrino charge and
magnetic form factors in the context of the standard model
supplied with $\mathrm{SU}(2)$-singlet right-handed neutrino.
Using the dimensional-regularization scheme, we start
(Section~\ref{NVF}) with the calculations of the one-loop Feynman
diagrams that contribute to the neutrino electromagnetic vertex
function $\Lambda_{\mu}(q)$ in the general $R_{\xi}$ gauge. It
should be noted that, contrary to the previous studies, we
explicitly account for non-vanishing neutrino mass. In
Section~\ref{f5FF} we examine the structure of the massive
neutrino electromagnetic vertex function. The decomposition of a
fermion vertex function in terms of the four well-known
electromagnetic form factors (presented, for instance, in
Refs.~\cite{Kim76,BegMarRud78}) has been established using general
principles such as the Lorentz and CP invariance and the
hermicity. We analyze this decomposition and verify it by means of
direct calculations in the case of massive neutrino within the
standard model supplied with $\mathrm{SU}(2)$-singlet right-handed
neutrino. Such direct calculations were never undertaken
previously.

We present the general expressions for the contributions to
neutrino electric form factor in Section~\ref{NCF}. Then in
Section~\ref{NEC} we study the neutrino electric charge and
analyze neutrino mass and gauge dependence of corresponding
contributions arising from different Feynman diagrams. Although
there is no doubt that the neutrino electric charge within the
standard model is a gauge independent and vanishing quantity,
however this fact has not been yet actually demonstrated in the
case of a massive neutrino. Our calculations allow us to determine
the neutrino mass and gauge parameter dependence of the one-loop
contributions to the neutrino charge. We also obtain a correct
gauge dependence for the contributions of several diagrams to the
neutrino charge that have been calculated in
Ref.~\cite{CabBerVidZep00} with mistakes. Within the one-loop
level we show that the neutrino electric charge is gauge
independent and vanishing in the "zeroth" and first order of the
expansion over the neutrino mass parameter
$b=\left(m_{\nu}/M_{W}\right)^{2}$. Moreover, for the particular
choice of the 't~Hooft-Feynman gauge we also demonstrate that the
neutrino charge is zero for arbitrary neutrino mass, i.e. in all
orders of the expansion over the parameter $b$. The obtained
formulae can be used for studying the massive neutrino charge
radius (Section~\ref{NCR}).

In Section~\ref{NMM} we consider the neutrino magnetic form factor
using the one-loop contributions to the neutrino electromagnetic
vertex derived in Section~\ref{NVF}. For each of the contributions
to the neutrino magnetic moment we derive the integral
representations that exactly account for the gauge-fixing
parameters as well as for the neutrino mass and corresponding
charged lepton mass parameters ($b$ and
$a=\left(m_{\ell}/M_{W}\right)^{2}$). Then for each of the
diagrams we perform an integration and obtain the explicitly
gauge-dependent contributions to the neutrino magnetic moment
accounting for the leading (zeroth) and next-to-leading (first)
order in the expansion over neutrino mass parameter $b$. The sum
of all the contributions turns out to be gauge parameter
independent. However, our results for several contributions to the
neutrino magnetic moment in the leading order in the neutrino mass
disagree with those of Ref.~\cite{CabBerVidZep00} for the
corresponding contributions. In particular, contrary to the
results of Ref.~\cite{CabBerVidZep00}, not all the contributions
are gauge independent. Our calculations enable one to reproduce
the correct value for the neutrino magnetic moment in any gauge
including also the unitary gauge for which the results of
Ref.~\cite{CabBerVidZep00} are incorrect. In this Section we get
final expressions for the massive neutrino magnetic moment in the
various ranges of neutrino, charged lepton and $W$ boson masses:
$m_\nu\ll m_\ell\ll M_W$, $m_\ell\ll m_\nu\ll M_W$, and $m_\ell\ll
M_W\ll m_\nu$. The last case amounts to a very heavy neutrino that
is not excluded by the LEP data (see, e.g., Ref.~\cite{Acc99}). We
also discuss the general formulae for the massive neutrino
magnetic form factor at non-zero momentum transfer.

The conclusions are made in Section~\ref{concl}. We also include a
list of the Feynman rules (Appendix~\ref{FeynRul}) and the typical
Feynman integrals (Appendix~\ref{FeynInt}) used in our
calculations.

\section{Vertex Function of Massive Neutrino\label{NVF}}

The matrix element of the electromagnetic current between neutrino
states can be presented in the form
\begin{equation}
  \langle{\nu}(p^{\prime})|J_{\mu}^\mathrm{EM}|\nu(p)\rangle=
  \bar{u}(p^{\prime})\Lambda_{\mu}(q)u(p),
\end{equation}
where the most general expression for the electromagnetic vertex
function $\Lambda_{\mu}(q)$ reads
\begin{equation}
  \label{J}
  \Lambda_{\mu}(q)=
  f_{Q}(q^{2})\gamma_{\mu}+f_{M}(q^{2})i\sigma_{\mu\nu}q^{\nu}-
  f_{E}(q^{2})\sigma_{\mu\nu}q^{\nu}\gamma_{5}+
  f_{A}(q^{2})(q^{2}\gamma_{\mu}-q_{\mu}{\not q})\gamma_{5}.
\end{equation}
Here $f_{Q}(q^{2})$, $f_{M}(q^{2})$, $f_{E}(q^{2})$ and
$f_{A}(q^{2})$ are respectively the electric, dipole electric,
dipole magnetic, and anapole neutrino form factors,
$q_{\mu}=p^{\prime}_{\mu}-p_{\mu}$,
$\sigma_{\mu\nu}=(i/2)[\gamma_{\mu},\gamma_{\nu}]$,
$\gamma_5=-i\gamma^0\gamma^1\gamma^2\gamma^3$. Their values at
$q^{2}=0$ determine the static electromagnetic properties of the
neutrino. In the case of Dirac neutrinos, which is considered in
this paper, the assumption of CP invariance combined with the
hermicity of the electromagnetic current $J_{\mu}^{EM}$ implies
that the electric dipole form factor vanishes. At zero momentum
transfer only $f_{Q}(0)$ and $f_{M}(0)$, which are called the
electric charge and the magnetic moment, respectively, contribute
to the Hamiltonian $H_\mathrm{int}\sim J_{\mu}^\mathrm{EM}A^{\mu}$
that describes the neutrino interaction with external
electromagnetic field $A^{\mu}$.

There is an important difference between the electromagnetic
vertex function representations in the cases of massive and
massless neutrino, respectively. If we consider a massless
particle, from Eq.~\eqref{J} it follows that the matrix element of
electromagnetic current can be expressed in terms of only one form
factor (see, for example, Ref.~\cite{Ros00})
\begin{equation}
  \bar{u}(p^{\prime})\Lambda_{\mu}(q)u(p)=
  f_{D}(q^{2})\bar{u}(p^{\prime})\gamma_{\mu}(1+\gamma_{5})u(p).
\end{equation}
Thus, the electric charge and anapole form factors are related to
the function $f_{D}(q^{2})$ by the trivial identities
\begin{equation}
  f_{Q}(q^{2})=f_{D}(q^{2}), \quad f_{A}(q^{2})=f_{D}(q^{2})/q^{2}.
\end{equation}
However in the case of a massive particle, there is no such simple
relation between the electric and anapole form factors since we
cannot neglect the $q_{\mu}{\not q}\gamma_{5}$-matrix term in the
anapole form factor. Moreover, the direct calculation of the
massive neutrino electromagnetic form factors shows that, besides
the ordinary electric charge and magnetic moment, each of the
Feynman diagrams gives non-zero contribution to the term
proportional to $\gamma_{\mu}\gamma_{5}$-matrix. These
contributions does not vanish even at $q^{2}=0$. This problem is
related to the decomposition of the massive neutrino
electromagnetic vertex function. Taking into account the
importance of this problem, we present the direct calculation that
verifies the decomposition given by Eq.~\eqref{J}. Using the
developed in the next section technique for studying the neutrino
electric charge, we find that the sum of contributions of the
complete set of Feynman diagrams to this additional additional
"form factor" is zero at $q^{2}=0$. The vanishing of the
considered "form factor" at $q^{2}\not=0$ for the particular
choice of the gauge is also demonstrated in the next subsection of
this paper.

\begin{figure}
  \subfigure[]
  {\label{prverta}
    \includegraphics{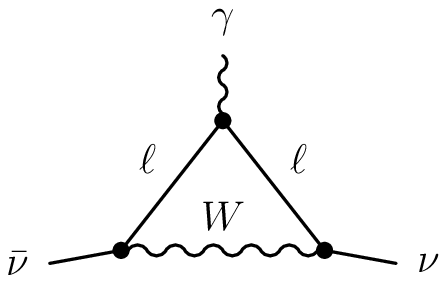}}
    \hspace{2cm}
  \subfigure[]
  {\label{prvertb}
  \includegraphics{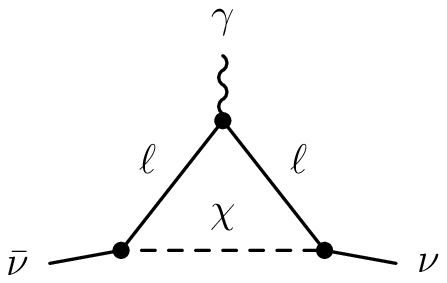}}
    \\
  \subfigure[]
  {\label{prvertc}
  \includegraphics{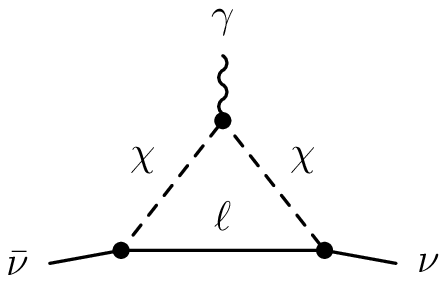}}
    \hspace{2cm}
  \subfigure[]
  {\label{prvertd}
  \includegraphics{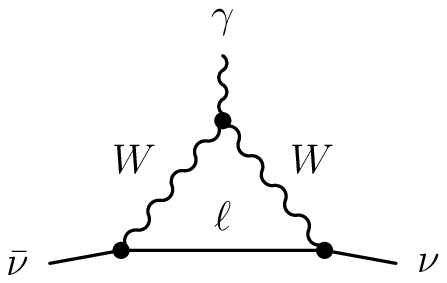}}
    \\
  \subfigure[]
  {\label{prverte}
  \includegraphics{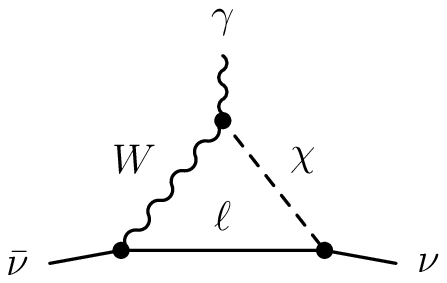}}
    \hspace{2cm}
  \subfigure[]
  {\label{prvertf}
  \includegraphics{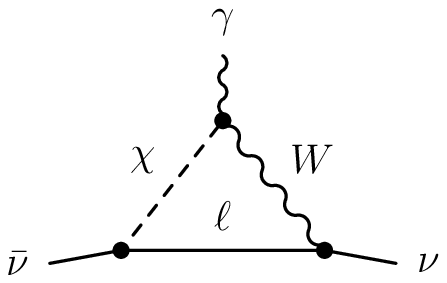}}
    \caption{\subref{prverta}-\subref{prvertf} proper vertices.}
\end{figure}

We present below the one-loop calculation of the electric charge
and magnetic moment of the massive neutrino within the context of
the standard model supplied with $\mathrm{SU}(2)$-singlet
right-handed neutrino in the general $R_{\xi}$ gauge. The one-loop
contributions to the neutrino electromagnetic vertex
$\Lambda_{\mu}(q)$ are given by the two types of Feynman diagrams:
the proper vertices (Fig.~\ref{prverta}-\ref{prvertf}) and the
$\gamma-Z$ self-energy diagrams
(Fig.~\ref{gZverta}-\ref{gZverth}). We use the Feynman rules given
in Appendix~\ref{FeynRul} to find the contributions to the
neutrino vertex function $\Lambda_{\mu}(q)$. In the dimensional-
regularization scheme the contributions of the proper vertices
diagrams (Fig.~\ref{prverta}-\ref{prvertf}) can be written as
\begin{equation}
  \label{L1}
  \Lambda^{(1)}_{\mu}=i{\frac{eg^{2}}{2}}
  \int {\frac{d^{N}k}{(2\pi)^{N}}}
  \left[
  g^{\kappa\lambda}-(1-\alpha)
  {\frac{k^{\kappa}k^{\lambda}}{k^{2}-\alpha M_{W}^{2}}}
  \right]
  {\frac{\gamma_{\kappa}^{L}
  (\not{p}^{\prime}-{\not k}+m_{\ell})\gamma_{\mu}
  ({\not p}-{\not k}+m_{\ell})\gamma_{\lambda}^{L}}
  {[(p^{\prime}-k)^{2}-m_{\ell}^{2}]
  [(p-k)^{2}-m_{\ell}^{2}][k^{2}-M_{W}^{2}]}},
\end{equation}
\begin{equation}
  \label{L2}
  \Lambda^{(2)}_{\mu}=i{\frac{eg^{2}}{2M_{W}^{2}}}
  \int {\frac{d^{N}k}{(2\pi)^{N}}}
  {\frac{(m_{\nu}P_{L}-m_{\ell}P_{R})
  ({\not p}^{\prime}-{\not k}+m_{\ell})\gamma_{\mu}
  ({\not p}-{\not k}+m_{\ell})
  (m_{\ell}P_{L}-m_{\nu}P_{R})}
  {[(p^{\prime}-k)^{2}-m_{\ell}^{2}]
  [(p-k)^{2}-m_{\ell}^{2}][k^{2}-\alpha M_{W}^{2}]}},
\end{equation}
\begin{equation}
  \label{L3}
  \Lambda^{(3)}_{\mu}=i{\frac{eg^{2}}{2M_{W}^{2}}}
  \int {\frac{d^{N}k}{(2\pi)^{N}}}
  (2k-p-p^{\prime})_{\mu}
  {\frac{(m_{\nu}P_{L}-m_{\ell}P_{R})
  ({\not k}+m_{\ell})
  (m_{\ell}P_{L}-m_{\nu}P_{R})}
  {[(p^{\prime}-k)^{2}-\alpha M_{W}^{2}]
  [(p-k)^{2}-\alpha M_{W}^{2}][k^{2}-m_{\ell}^{2}]}},
\end{equation}
\begin{multline}
  \label{L4}
  \Lambda^{(4)}_{\mu}=i{\frac{eg^{2}}{2}}
  \int {\frac{d^{N}k}{(2\pi)^{N}}}
  \gamma_{\kappa}^{L}({\not k}+
  m_{\ell})\gamma_{\lambda}^{L}
  \left[
  \delta^{\kappa}_{\beta}-(1-\alpha)
  {\frac{(p^{\prime}-k)^{\kappa}(p^{\prime}-k)_{\beta}}
  {(p^{\prime}-k)^{2}-\alpha M_{W}^{2}}}
  \right]\times
  \\
  \left[
  \delta^{\lambda}_{\gamma}-(1-\alpha)
  {\frac{(p-k)^{\lambda}(p-k)_{\gamma}}
  {(p-k)^{2}-\alpha M_{W}^{2}}}
  \right]\times
  \\
  {\frac{\delta^{\beta}_{\mu}(2p^{\prime}-p-k)^{\gamma}+
  g^{\beta\gamma}(2k-p-p^{\prime})_{\mu}+
  \delta^{\gamma}_{\mu}(2p-p^{\prime}-k)^{\beta}}
  {[(p^{\prime}-k)^{2}-M_{W}^{2}]
  [(p-k)^{2}-M_{W}^{2}][k^{2}-m_{\ell}^{2}]}},
\end{multline}
\begin{multline}
  \label{L56}
  \Lambda^{(5)+(6)}_{\mu}=i{\frac{eg^{2}}{2}}
  \int
  {\frac{d^{N}k}{(2\pi)^{N}}}\times
  \\
  \Big\{
  {\frac{\gamma_{\beta}^{L}({\not k}-m_{\ell})
  (m_{\ell}P_{L}-m_{\nu}P_{R})}
  {[(p^{\prime}-k)^{2}-M_{W}^{2}]
  [(p-k)^{2}-\alpha M_{W}^{2}][k^{2}-m_{\ell}^{2}]}}
  \left[
  \delta^{\beta}_{\mu}-(1-\alpha)
  {\frac{(p^{\prime}-k)^{\beta}(p^{\prime}-k)_{\mu}}
  {(p^{\prime}-k)^{2}-\alpha M_{W}^{2}}}
  \right]-
  \\
  {\frac{(m_{\nu}P_{L}-m_{\ell}P_{R})
  ({\not k}-m_{\ell})\gamma_{\beta}^{L}}
  {[(p^{\prime}-k)^{2}-\alpha M_{W}^{2}]
  [(p-k)^{2}-M_{W}^{2}][k^{2}-m_{\ell}^{2}]}}
  \left[
  \delta^{\beta}_{\mu}-(1-\alpha)
  {\frac{(p-k)^{\beta}(p-k)_{\mu}}
  {(p-k)^{2}-\alpha M_{W}^{2}}}
  \right]
  \Big\},
\end{multline}
where  $m_{\nu}$, $M_{W}$ and $m_{\ell}$ are the masses of
neutrino, $W$ boson, and the charged isodoublet partner of the
neutrino, respectively, $e$ is the proton charge, $g$ is the
coupling constant of the standard model, $\theta_{W}$ is the
Weinberg angle, $\alpha=1/\xi$ is the gauge parameter of $W$
boson, $P_{L,R}=(1\pm\gamma_{5})/2$ are the projection operators.
\begin{figure}
  \subfigure[]
  {\label{gZverta}
    \includegraphics{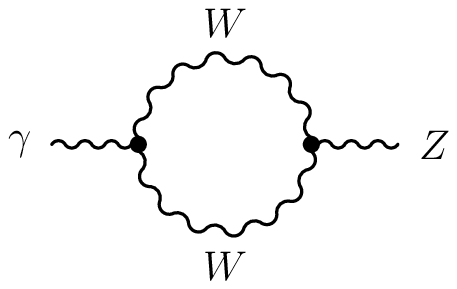}}
    \hspace{2cm}
  \subfigure[]
  {\label{gZvertb}
  \includegraphics{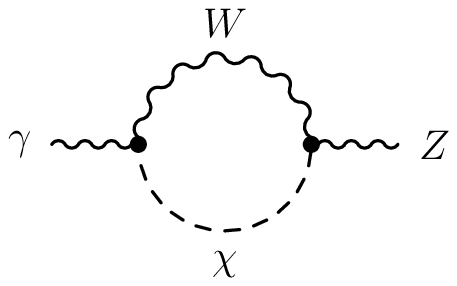}}
    \\
  \subfigure[]
  {\label{gZvertc}
  \includegraphics{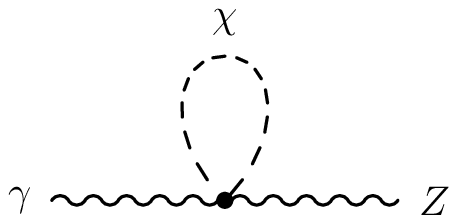}}
    \hspace{2cm}
  \subfigure[]
  {\label{gZvertd}
  \includegraphics{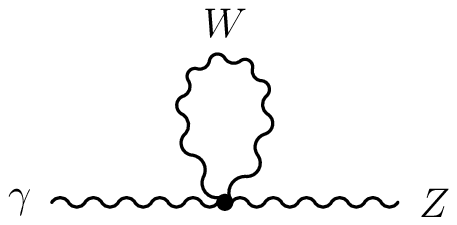}}
    \\
  \subfigure[]
  {\label{gZverte}
  \includegraphics{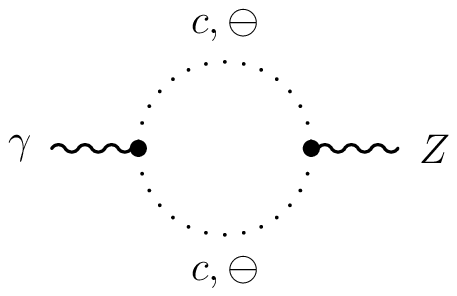}}
    \hspace{2cm}
  \subfigure[]
  {\label{gZvertf}
  \includegraphics{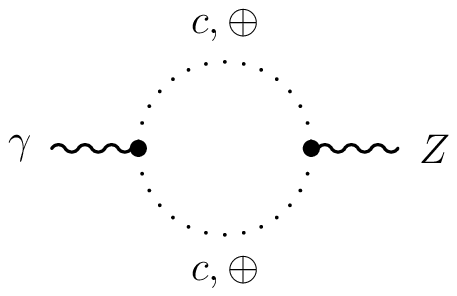}}
    \\
  \subfigure[]
  {\label{gZvertg}
  \includegraphics{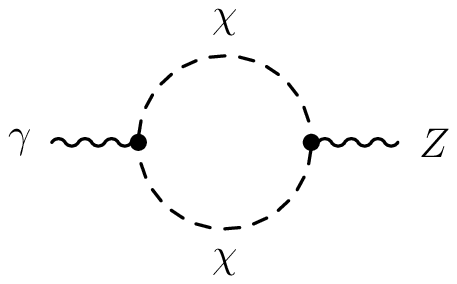}}
    \hspace{2cm}
  \subfigure[]
  {\label{gZverth}
  \includegraphics{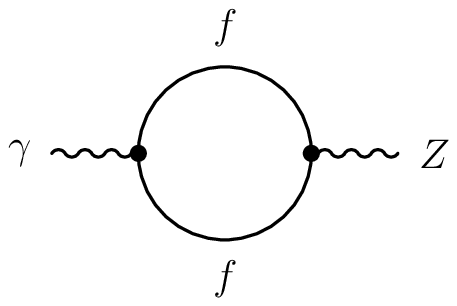}}
    \caption{\subref{gZverta}-\subref{gZverth}
    The $\gamma -Z$ self-energy diagram.
    $f$ denotes the electron, muon, and $\tau$-lepton as well as $u$, $c$ ,$t$,
    $d$, $s$, and $b$ quarks.}
\end{figure}

The contributions of $\gamma -Z$ self-energy diagrams
(Fig.~\ref{gZverta}-\ref{gZverth}) to the vertex
$\Lambda_{\mu}(q)$ are given by
\begin{equation}
  \label{L7-14}
  \Lambda_{\mu}^{(j)}(q)={\frac{g}{2\cos\theta_{W}}}
  \Pi_{\mu\nu}^{(j)}(q){\frac{1}{q^{2}-M_{Z}^{2}}} \left\{
  g^{\nu\alpha}-(1-\alpha_{Z})
  {\frac{q^{\nu}q^{\alpha}}{q^{2}-\alpha_{Z}M_{Z}^{2}}} \right\}
  \gamma_{\alpha}^{L}, \quad  j=7,\dots,14,
\end{equation}
where
\begin{multline}
  \label{P7}
  \Pi^{(7)}_{\mu\nu}(q)=-ieg\cos \theta_{W}
  \int {\frac{d^{N}k}{(2\pi)^{N}}}
  {\frac{1}{[(k-q)^{2}-M_{W}^{2}][k^{2}-M_{W}^{2}]}}\times
  \\
  \left[
  g_{\gamma\alpha}-(1-\alpha)
  {\frac{(k-q)_{\gamma}(k-q)_{\alpha}}{(k-q)^{2}-\alpha M_{W}^{2}}}
  \right]
  \left[
  g_{\beta\lambda}-(1-\alpha)
  {\frac{k_{\beta}k_{\lambda}}{k^{2}-\alpha M_{W}^{2}}}
  \right]\times
  \\
  \left[
  (k+q)^{\gamma}\delta^{\beta}_{\mu}+(q-2k)_{\mu}g^{\beta\gamma}+
  (k-2q)^{\beta}\delta^{\gamma}_{\mu}
  \right]
  \left[
  (k+q)^{\alpha}\delta^{\lambda}_{\nu}+(q-2k)_{\nu}g^{\alpha\lambda}+
  (k-2q)^{\lambda}\delta^{\alpha}_{\nu}
  \right],
\end{multline}
\begin{equation}
  \label{P8}
  \Pi^{(8)}_{\mu\nu}(q)=
  -2ieg{\frac{\sin^{2}\theta_{W}}{\cos \theta_{W}}}
  M_{W}^{2}
  \int {\frac{d^{N}k}{(2\pi)^{N}}}
  {\frac{1}{[(k-q)^{2}-\alpha M_{W}^{2}][k^{2}-M_{W}^{2}]}}
  \left[
  g_{\mu\nu}-(1-\alpha)
  {\frac{k_{\mu}k_{\nu}}{k^{2}-\alpha M_{W}^{2}}}
  \right],
\end{equation}
\begin{equation}
  \label{P9}
  \Pi^{(9)}_{\mu\nu}(q)=ieg
  {\frac{\cos^{2}\theta_{W}-\sin^{2}\theta_{W}}{\cos \theta_{W}}}
  \int {\frac{d^{N}k}{(2\pi)^{N}}}
  {\frac{g_{\mu\nu}}{k^{2}-\alpha M_{W}^{2}}},
\end{equation}
\begin{equation}
  \label{P10}
  \Pi^{(10)}_{\mu\nu}(q)=-ieg\cos \theta_{W}
  \int {\frac{d^{N}k}{(2\pi)^{N}}}
  {\frac{\delta^{\alpha}_{\mu}\delta^{\beta}_{\nu}+
  \delta^{\beta}_{\mu}\delta^{\alpha}_{\nu}-
  2g^{\alpha\beta}g_{\mu\nu}}
  {[k^{2}-M_{W}^{2}]}}
  \left[
  g_{\alpha\beta}-(1-\alpha)
  {\frac{k_{\alpha}k_{\beta}}{k^{2}-\alpha M_{W}^{2}}}
  \right],
\end{equation}
\begin{equation}
  \label{P1112}
  \Pi^{(11)+(12)}_{\mu\nu}(q)=2ieg\cos \theta_{W} \int
  {\frac{d^{N}k}{(2\pi)^{N}}} {\frac{k_{\mu}(k-q)_{\nu}}
  {[(k-q)^{2}-\alpha M_{W}^{2}][k^{2}-\alpha M_{W}^{2}]}},
\end{equation}
\begin{equation}
  \label{P13}
  \Pi^{(13)}_{\mu\nu}(q)=ieg
  {\frac{\sin^{2}\theta_{W}-\cos^{2}\theta_{W}}{2\cos \theta_{W}}}
  \int {\frac{d^{N}k}{(2\pi)^{N}}}
  (2k-q)_{\mu}(2k-q)_{\nu}
  {\frac{1}{[(k-q)^{2}-\alpha M_{W}^{2}][k^{2}-\alpha M_{W}^{2}]}},
\end{equation}
\begin{multline}
  \label{P14}
  \Pi^{(14)}_{\mu\nu}(q)= {\frac{ i e g}{2\cos
  \theta_{W}}}\sum_{f}^{}Q_{f} \int {\frac{d^{N}k}{(2\pi)^{N}}}
  {\frac{1}{[(k-q)^{2}-m_{f}^{2}][k^{2}-m_{f}^{2}]}}\times
  \\
  \mathrm{Tr}
  \left[
  \gamma_{\mu}({\not k}+m_{f})\gamma_{\nu}
  \left\{
  \pm{\frac{1}{2}}-2Q_{f}\sin^{2}\theta_{W}
  \pm{\frac{1}{2}}\gamma_{5}
  \right\}
  ({\not k}-{\not q}+m_{f})
  \right].
\end{multline}
Here $M_{Z}$ and $\alpha_{Z}$ are respectively the mass and gauge
parameter of $Z$ boson. In Eq.~\eqref{P14}, "$-$" and "$+$" stand
for the "upper" ($u$, $c$, and $t$ quarks) and "lower" (electron,
muon, $\tau$-lepton as well as $d$, $s$, and $b$ quarks)
components of an isodoublet, $m_{f}$ and $Q_{f}$ are the mass and
electric charge (in the units of $e$) of a fermion circulating
within the loop.

It is convenient to decompose each of the $\gamma-Z$ self-energy
contributions at arbitrary $q^{2}$ and explicitly extract the
transversal term:
\begin{equation}
  \label{P7-14}
  \Pi_{\mu\nu}^{(j)}(q)=A^{(j)}(\alpha,q^{2}) \left(
  g_{\mu\nu}-{\frac{q_{\mu}q_{\nu}}{q^{2}}} \right)+
  B^{(j)}(\alpha,q^{2})g_{\mu\nu}, \quad
  j=7,\dots,14.
\end{equation}
Using Eqs.~\eqref{P7}-\eqref{P14} for the contributions of the
$\gamma-Z$ self-energy diagrams in the form of the Feynman
integrals as well as Eq.~\eqref{P7-14}, it is possible to present
the functions $A^{(j)}(\alpha,q^2)$ and $B^{(j)}(\alpha,q^2)$
($j=7,\dots,14$) in the explicit form
\begin{multline}
  \label{A7}
  A^{(7)}(\alpha,q^2)=
  2\cos^{3}\theta_{W}\sin\theta_{W}M_{W}^{2}M_{Z}^{2}
  \widetilde{G}_{F}
  \tau
  \Big[
  \omega\left(-\frac{14}{3}+\alpha\right)+
  \frac{1}{6}+\frac{\alpha}{2}-
  \\
  2\tau\int_{0}^{1}dx (1-x^{2})^{2}
  \left\{\ln(1-\zeta-x(1-\alpha))-\ln(1-\zeta)\right\}+
  2\int_{0}^{1}dx (5x^{2}-5x-1)\ln(1-\zeta)-
  \\
  2\int_{0}^{1}dx (4x^{2}-3)
  \left\{(1-\zeta-x(1-\alpha))\ln(1-\zeta-x(1-\alpha))-
  (1-\zeta)\ln(1-\zeta)\right\}+
  \\
  \frac{\tau}{2}\int_{0}^{1}dx
  \{2(1-\zeta-x(1-\alpha))\ln(1-\zeta-x(1-\alpha))-
  (1-\zeta)\ln(1-\zeta)-
  (\alpha-\zeta)\ln(\alpha-\zeta)\}
  \Big],
\end{multline}
\begin{equation}
  \label{A8}
  A^{(8)}(\alpha,q^2)=
  -4\cos\theta_{W}\sin^{3}\theta_{W}M_{W}^{2}M_{Z}^{2}
  \widetilde{G}_{F}
  \tau
  \int_{0}^{1}dx\thinspace x^{2}
  \left\{\ln(1-\zeta-x(1-\alpha))-
  \ln(\alpha-\zeta)\right\},
\end{equation}
\begin{equation}
  \label{A9}
  A^{(9)}(\alpha,q^2)=0,
\end{equation}
\begin{equation}
  \label{A10}
  A^{(10)}(\alpha,q^2)=0,
\end{equation}
\begin{equation}
  \label{A1112}
  A^{(11)+(12)}(\alpha,q^2)=
  2\cos^{3}\theta_{W}\sin\theta_{W}M_{W}^{2}M_{Z}^{2}
  \widetilde{G}_{F}
  \tau
  \Big[
  \frac{\omega}{3}+
  2\int_{0}^{1}dx\thinspace x(1-x)
  \ln(\alpha-\zeta)
  \Big],
\end{equation}
\begin{equation}
  \label{A13}
  A^{(13)}(\alpha,q^2)=
  (\sin^{2}\theta_{W}-\cos^{2}\theta_{W})
  \cos\theta_{W}\sin\theta_{W}M_{W}^{2}M_{Z}^{2}
  \widetilde{G}_{F}
  \tau
  \Big[
  -\frac{\omega}{3}-
  \int_{0}^{1}dx (2x-1)^{2}
  \ln(\alpha-\zeta)
  \Big],
\end{equation}
\begin{multline}
  \label{A14}
  A^{(14)}(\alpha,q^2)=
  8\cos\theta_{W}\sin\theta_{W}M_{W}^{2}M_{Z}^{2}
  \widetilde{G}_{F}
  \tau
  \Big[
  \frac{\omega}{6}
  \left(-3-\frac{28}{3}\sin^{2}\theta_{W}\right)+
  \\
  \sum_{f}Q_{f}
  \left(\pm\frac{1}{2}-2Q_{f}\sin^{2}\theta_{W}\right)
  \left\{\frac{1}{6}\ln\left(\frac{m_{f}}{M}\right)^{2}+
  \int_{0}^{1}dx\thinspace x(1-x)
  \ln\left(1-\right(M/m_{f}\left)^{2}\zeta\right)
  \right\}
  \Big],
\end{multline}
\begin{multline}
  \label{B7}
  B^{(7)}(\alpha,q^2)=
  2\cos^{3}\theta_{W}\sin\theta_{W}M_{W}^{2}M_{Z}^{2}
  \widetilde{G}_{F}
  \Big[
  \omega\left(\frac{\tau}{2}-
  \frac{12+3\alpha(1+\alpha)}{2}\right)+
  \frac{3}{4}(2+\alpha(1+\alpha))-
  \frac{\tau}{24}(25+3\alpha)-
  \\
  3\tau
  \int_{0}^{1}dx (2x-1)^{2}\ln(1-\zeta)-
  9\int_{0}^{1}dx (1-\zeta)\ln(1-\zeta)-
  \\
  3\tau
  \int_{0}^{1}dx\thinspace x^{2}
  \left\{(1-\zeta-x(1-\alpha))\ln(1-\zeta-x(1-\alpha))-
  (1-\zeta)\ln(1-\zeta)\right\}-
  \\
  \frac{9}{2}\int_{0}^{1}dx
  \left\{(1-\zeta-x(1-\alpha))^{2}\ln(1-\zeta-x(1-\alpha))-
  (1-\zeta)^{2}\ln(1-\zeta)\right\}
  \Big],
\end{multline}
\begin{multline}
  \label{B8}
  B^{(8)}(\alpha,q^2)=
  2\cos\theta_{W}\sin^{3}\theta_{W}M_{W}^{2}M_{Z}^{2}
  \widetilde{G}_{F}
  \Big[
  -\omega\frac{3+\alpha}{2}-
  \frac{1-\alpha}{2}-
  2\int_{0}^{1}dx \ln(1-\zeta-x(1-\alpha))+
  \\
  \int_{0}^{1}dx
  \left\{(1-\zeta-x(1-\alpha))\ln(1-\zeta-x(1-\alpha))-
  (\alpha-\zeta)\ln(\alpha-\zeta)\right\}+
  \\
  2\tau\int_{0}^{1}dx\thinspace x^{2}
  \left\{\ln(1-\zeta-x(1-\alpha))-
  \ln(\alpha-\zeta)\right\}
  \Big],
\end{multline}
\begin{equation}
  \label{B9}
  B^{(9)}(\alpha,q^2)=
  2(\cos^{2}\theta_{W}-\sin^{2}\theta_{W})
  \cos\theta_{W}\sin\theta_{W}M_{W}^{2}M_{Z}^{2}
  \widetilde{G}_{F}
  \Big[
  \alpha(\omega-1)+\alpha\ln\alpha
  \Big],
\end{equation}
\begin{equation}
  \label{B10}
  B^{(10)}(\alpha,q^2)=
  6\cos^{3}\theta_{W}\sin\theta_{W}M_{W}^{2}M_{Z}^{2}
  \widetilde{G}_{F}
  \Big[
  \omega\frac{3+\alpha^{2}}{2}-\frac{1}{4}-
  \frac{5\alpha^{2}}{12}+
  \frac{\alpha^{2}\ln\alpha}{2}
  \Big],
\end{equation}
\begin{multline}
  \label{B1112}
  B^{(11)+(12)}(\alpha,q^2)=
  2\cos^{3}\theta_{W}\sin\theta_{W}M_{W}^{2}M_{Z}^{2}
  \widetilde{G}_{F}
  \Big[
  \omega\left(\alpha-\frac{\tau}{2}\right)-
  \alpha+\frac{\tau}{6}+
  \\
  \int_{0}^{1}dx
  (\alpha-\zeta)\ln(\alpha-\zeta)-
  2\tau\int_{0}^{1}dx\thinspace x(1-x)
  \ln(\alpha-\zeta)
  \Big],
\end{multline}
\begin{multline}
  \label{B13}
  B^{(13)}(\alpha,q^2)=
  2(\sin^{2}\theta_{W}-\cos^{2}\theta_{W})
  \cos\theta_{W}\sin\theta_{W}M_{W}^{2}M_{Z}^{2}
  \widetilde{G}_{F}
  \Big[
  \alpha(\omega-1)+\frac{\tau}{6}+
  \\
  \int_{0}^{1}dx
  (\alpha-\zeta)\ln(\alpha-\zeta)+
  \frac{\tau}{2}\int_{0}^{1}dx (2x-1)^{2}
  \ln(\alpha-\zeta)
  \Big],
\end{multline}
\begin{equation}
  \label{B14}
  B^{(14)}(\alpha,q^2)=0,
\end{equation}
where
\begin{equation*}
  \widetilde{G}_{F}=\frac{G_{F}}{4\pi^{2}\sqrt{2}},
  \quad
  \omega=-{\frac{1}{\varepsilon}}-\ln (4\pi^{2})+{\mathbb C}-
  \ln
  \left(
  {\frac{\lambda^{2}}{M_{W}^{2}}}
  \right),
\end{equation*}
and $G_{F}$ is the Fermi constant, $\zeta=x(1-x)\tau$,
$\tau=q^{2}/M_{W}^{2}$.

In the derivation of Eqs.~\eqref{A7}-\eqref{B14} we have used the
properties of the $\gamma$ matrix algebra in $N$ dimensional space
and the expressions for the characteristic loop integrals
presented in Appendix~\ref{FeynInt}.

\subsection{\label{f5FF} The decomposition of massive neutrino
electromagnetic vertex function}

In the direct calculations of the massive neutrino electromagnetic
vertex function, taking into accounting the complete set of the
Feynman diagrams, one reveals that, besides the well-known four
terms, in Eq.~\eqref{J} appears an additional term proportional to
$\gamma_{\mu}\gamma_{5}$-matrix. Therefore we introduce the
additional "form factor" $f_{5}(q^{2})$. In this subsection we
analyze this "form factor" and  show by explicit calculations that
$f_{5}(q^{2})=0$ for arbitrary $q^{2}$ and for particular choices
of the particles' masses and gauge-fixing parameters.

Let us first consider the value of $f_{5}(q^{2})$ at $q^{2}=0$:
$\varphi=f_{5}(q^{2}=0)$. The contributions to the "charge"
$\varphi$ of the proper vertices diagrams
(Fig.~\ref{prverta}-\ref{prvertf}) have the form:
\begin{multline}
  \label{phi1}
  \varphi^{(1)}(a,b,\alpha)=
  {\frac{eG_{F}}{4\pi^{2}\sqrt{2}}}M_{W}^{2}
  \bigg\{
  \omega{\frac{\alpha}{2}}+1+{\frac{1-\alpha}{12}}+
  \int_{0}^{1} dz (1-z) \ln D-
  \\
  \int_{0}^{1} dz (1-z) (a-bz^{2}) {\frac{1}{D}}+
  {\frac{b}{2}}\int_{0}^{1} dz (1-z)^{3}(a-bz^{2})
  \left[
  {\frac{1}{D}_{\alpha}}-{\frac{1}{D}}
  \right]-
  \\
  {\frac{1}{2}}\int_{0}^{1} dz (1-z) (a-b+6bz(1-z))
  \left[
  \ln D_{\alpha}-\ln D
  \right]+
  3\int_{0}^{1} dz (1-z)
  \left[
  D_{\alpha}\ln D_{\alpha}-D\ln D
  \right]
  \bigg\},
\end{multline}
\begin{multline}
  \label{phi2}
  \varphi^{(2)}(a,b,\alpha)=
  {\frac{eG_{F}}{4\pi^{2}\sqrt{2}}}M_{W}^{2}
  \bigg\{
  {\frac{a-b}{2}}
  \left(
  {\frac{\omega}{2}}+{\frac{1}{2}}+
  \int_{0}^{1} dz (1-z) \ln D_{\alpha}
  \right)-
  \\
  {\frac{1}{2}}\int_{0}^{1} dz (1-z)
  (a^{2}-abz^{2}+b^{2}z^{2}-ab) {\frac{1}{D_{\alpha}}}
  \bigg\},
\end{multline}
\begin{equation}
  \label{phi3}
  \varphi^{(3)}(a,b,\alpha)=
  {\frac{eG_{F}}{4\pi^{2}\sqrt{2}}}M_{W}^{2}
  {\frac{a-b}{2}}
  \left(
  -{\frac{\omega}{2}} -
  \int_{0}^{1} dz\thinspace z \ln D_{\alpha}
  \right),
\end{equation}
\begin{multline}
  \label{phi4}
  \varphi^{(4)}(a,b,\alpha)=
  {\frac{eG_{F}}{4\pi^{2}\sqrt{2}}}M_{W}^{2}
  \bigg\{
  -\omega{\frac{3}{4}}(1+\alpha)-1-
  3\int_{0}^{1} dz\thinspace z \ln D +
  b\int_{0}^{1} dz\thinspace z^{2}(1-z) {\frac{1}{D}}-
  \\
  {\frac{9}{2}}\int_{0}^{1} dz \int_{0}^{z} dy
  \left[
  (D_{\alpha}+y(1-\alpha))\ln (D_{\alpha}+y(1-\alpha))- D\ln D
  \right]-
  \\
  b^{2}\int_{0}^{1} dz \int_{0}^{z} dy (1-z)^{2}(z(1-z)-2y)
  \left[
  {\frac{1}{D_{\alpha}+y(1-\alpha))}}-{\frac{1}{D}}
  \right]-
  \\
  {\frac{b}{2}}
  \int_{0}^{1} dz \int_{0}^{z} dy (7-18z+11z^{2})
  \left[
  \ln (D_{\alpha}+y(1-\alpha))-\ln D
  \right]
  \bigg\},
\end{multline}
\begin{multline}
  \label{phi56}
  \varphi^{(5)+(6)}(a,b,\alpha)=
  {\frac{eG_{F}}{4\pi^{2}\sqrt{2}}}M_{W}^{2}
  \bigg\{
  \int_{0}^{1} dz \int_{0}^{z} dy (a-bz)
  {\frac{1}{D_{\alpha}+y(1-\alpha)}}-
  \\
  {\frac{a-b}{2}}\int_{0}^{1} dz \int_{0}^{z} dy
  \left[
  \ln (D_{\alpha}+y(1-\alpha))-\ln D_{\alpha}
  \right]
  \bigg\},
\end{multline}
where
\begin{equation}
  a=
  \left(
  \frac{m_\ell}{M_W}
  \right)^2
  \thickspace
  \text{and}
  \thickspace
  b=
  \left(
  \frac{m_\nu}{M_W}
  \right)^2
\end{equation}
are the charged lepton and neutrino mass parameters, respectively,
$D_{\alpha}=a+(\alpha-a)z-bz(1-z)$ and
$D=D_{\alpha=1}=a+(1-a)z-bz(1-z)$.

In Eqs.~\eqref{phi1}-\eqref{phi56} we assume the mass parameters
$a$ and $b$ as well as the gauge parameter $\alpha$ to be
arbitrary. We have calculated the integrals in
Eqs.~\eqref{phi1}-\eqref{phi56}, however the results, being
expressed in elementary functions, are rather cumbersome.
Therefore we also perform corresponding integrations in the first
two terms of expansion over the neutrino mass parameter $b$ for
arbitrary values of the charged lepton mass parameter $a$ and the
gauge-fixing parameter $\alpha$. In this case the sum of the
proper vertices diagrams contributed to the "charge" $\varphi$ can
be written as
\begin{equation}
  \label{phi1-6}
  \varphi^{(\mathrm{prop.vert.})}(a,b,\alpha)=
  {\frac{eG_{F}}{4\pi^{2}\sqrt{2}}}M_{W}^{2}
  \sum_{i=1}^{6} \{ \bar{\varphi}^{(i)}_{0}(a,\alpha)+
  b\bar{\varphi}^{(i)}_{1}(a,\alpha)+
  \mathcal{O}(b^{2}) \}.
\end{equation}

The contributions of the $\gamma-Z$ self-energy diagrams
(Fig.~\ref{gZverta}-\ref{gZverth}) to the "charge" $\varphi$
coincides with those to the neutrino electric charge
$Q^{(\gamma-Z)}$ and thus are given by Eq.~\eqref{Q7-14} (the
details of the neutrino electric charge calculations can be found
in Section~\ref{NEC}). We have calculated the functions
$\bar{\varphi}^{(i)}_{0}(a,\alpha)$ and
$\bar{\varphi}^{(i)}_{1}(a,\alpha)$ and found out that the sum of
all the contributions $\bar{\varphi}^{(i)}_{0}(a,\alpha)$ exactly
cancels the contribution of the $\gamma-Z$ self-energy diagrams.
This our result corresponds to the case of a massless particle.
Therefore the "charge" $\varphi$ of a massless neutrino is zero.
Then, summing the contributions
$\bar{\varphi}^{(i)}_{1}(a,\alpha)$ we reveal that the value of
the "charge" $\varphi$ is also zero in the next order of the
expansion over the neutrino mass parameter $b$.

Now let us consider the value of the "form factor" $f_{5}(q^{2})$
at non-zero momentum transfer. In the subsequent calculations we
have to fix the gauge in order to simplify the formulae. We set
$\alpha_{Z}=\infty$ and $\alpha=1$ that corresponds to the unitary
gauge for $Z$ boson and the 't~Hooft-Feynman gauge for $W$ boson.
However, we do not restrict ourselves considering either light
neutrino or light charged lepton: the mass parameters $a$ and $b$
are arbitrary in all our calculations. In this case the function
$B(q^{2})$ in the decomposition of the $\gamma-Z$ self-energy
diagrams [see Eq.~\eqref{P7-14}] takes on the form
\begin{equation}
  \label{Bq2gen}
  B(q^{2})=\sum_{j=7}^{14}B^{(j)}(q^{2})=
  2\cos\theta_{W}\sin\theta_{W}M_{W}^{2}M_{Z}^{2}
  \widetilde{G}_{F}
  (-2\omega+
  g_{c}(\tau)\cos^{2}\theta_{W}+g_{s}(\tau)\sin^{2}\theta_{W}),
\end{equation}
where
\begin{align*}
  g_{c}(\tau) &=-\frac{7}{6}\tau-
  \frac{7\tau}{2}\int_{0}^{1}dx(6x^{2}-6x+1)\ln (1-\zeta)-
  9\int_{0}^{1}dx\ln (1-\zeta),
  \\
  g_{s}(\tau) &=\frac{\tau}{6}+
  \frac{\tau}{2}\int_{0}^{1}dx(6x^{2}-6x+1)\ln (1-\zeta)-
  \int_{0}^{1}dx\ln (1-\zeta).
\end{align*}
We note that $g_{s}(0)=g_{c}(0)=0$. To show that
$g_{s}(\tau)=g_{c}(\tau)$ for any value of $\tau$ we consider
their difference $g(\tau)=g_{s}(\tau)-g_{c}(\tau)$. The function
$g(\tau)$ can be represented as follows
\begin{equation*}
  g(\tau)=\frac{4}{3}\tau+
  8\int_{0}^{1}dx\ln (1-\zeta)+
  4\tau\int_{0}^{1}dx(6x^{2}-6x+1)\ln (1-\zeta).
\end{equation*}
Expanding $\ln (1-\zeta)$ in the formal series
\begin{equation*}
  \ln (1-\zeta)=-\sum_{k=0}^{\infty}
  \frac{\tau^{k}}{k}x^{k}(1-x)^{k},
\end{equation*}
and performing the integration with the help of the formula
\begin{equation}
  \label{BETAF}
  \int_{0}^{1}dx\thinspace x^{l}(1-x)^{s}=
  \frac{(l+1)!(s+1)!}{(l+s+2)!},
\end{equation}
we receive that $g(\tau)=0$ for any value of $\tau$. Thus, the
expression for $B(q^{2})$ can be represented in the form
\begin{equation*}
  B(q^{2})=
  2\cos\theta_{W}\sin\theta_{W}M_{W}^{2}M_{Z}^{2}
  \widetilde{G}_{F}
  (-2\omega+g_{s}(\tau)).
\end{equation*}
It is interesting to note that, in contrast to Eq.~\eqref{Bq2gen},
in this expression the dependence on $\cos^2\theta_{W}$ and
$\sin^2\theta_{W}$ is absent. We have also verified that the
analogous property of the function $B(q^2)$ remains valid in the
case of an arbitrary gauge.

The $\gamma-Z$ self-energy diagrams contribute to the
$f_{5}(q^{2})$ "form factor" according to the formula
\begin{equation*}
  f_{5}^{(\gamma-Z)}(q^{2})=
  -\frac{g}{4M_{Z}^{2}\cos^{2}\theta_{W}}B(q^{2}).
\end{equation*}

Now we turn to the contributions of the proper vertices diagrams
(shown in Fig.~\ref{prverta}-\ref{prvertf}) to the $f_{5}(q^{2})$
"form factor" at arbitrary $q^{2}$:
\begin{multline}
  \label{f5pv}
  f_{5}^{(\mathrm{prop. vert.})}(q^{2})=
  \sum_{i=1}^{6}f_{5}^{(i)}(q^{2})=
  \\
  \frac{eG_{F}}{4\pi^{2}\sqrt{2}}M_{W}^{2}
  \bigg\{
  -\omega+
  \int_{0}^{1}dz\int_{0}^{z}dy\ln (D^{\prime}-\tau y(z-y))-
  3\int_{0}^{1}dz\int_{0}^{z}dy\ln (D-\tau y(z-y))+
  \\
  \frac{a-b}{2}
  \left(
  \int_{0}^{1}dz\int_{0}^{z}dy\ln (D^{\prime}-\tau y(z-y))-
  \int_{0}^{1}dz\int_{0}^{z}dy\ln (D-\tau y(z-y))+
  \frac{1}{2}
  \right)+
  \\
  \left(
  1+\frac{a-b}{2}
  \right)
  \int_{0}^{1}dz\int_{0}^{z}dy
  \frac{1}{D^{\prime}-\tau y(z-y)}
  \bigg[
  -(a-b(1-z)^{2})+
  \tau\left(
  \frac{1}{2}z(1-z)+
  y(z-y)
  \right)
  \bigg]+
  \\
  \int_{0}^{1}dz\int_{0}^{z}dy
  \frac{1}{D-\tau y(z-y)}\times
  \\
  \bigg[
  a-bz^{2}+
  \tau\left(
  3y(z-y)-\frac{1}{4}z-\frac{1}{2}z^{2}+\frac{1}{2}y-
  \frac{1}{4}(a-b)z^{2}+(a-b)y(z-y)
  \right)
  \bigg]
  \bigg\},
\end{multline}
where $D^{\prime}=1+(a-1)z-bz(1-z)$. Eq.~\eqref{f5pv} can be
analyzed in the same manner as we have treated the function
$g(\tau)$. For instance, let us present the calculations of one of
the integrals in Eq.~\eqref{f5pv}
\begin{equation}
  \label{IT}
  I(\tau)=\int_{0}^{1}dz\int_{0}^{z}dy
  (a-bz^{2})
  \frac{1}{D-\tau y(z-y)}.
\end{equation}
We again expand the integrand in Eq.~\eqref{IT} in the formal
series
\begin{equation*}
  \frac{1}{D-\tau y(z-y)}=\frac{1}{D}
  \sum_{k=0}^{\infty}
  \left(
  \frac{\tau}{D}
  \right)^{k}
  y^{k}(z-y)^{k}.
\end{equation*}
Then, we carry out the integration over the variable $y$ using
Eq.~\eqref{BETAF}. The obtained result should be transformed
according to the identity
\begin{equation*}
  \int_{0}^{1}dz\frac{z^{l}}{D^{k+1}}(a-bz^{2})=
  \frac{1}{k}+\frac{k-l-1}{k}
  \int_{0}^{1}dz\frac{z^{l}}{D^{k}},
  \quad k\geq 1, \quad l\geq 0,
\end{equation*}
which can be proven by means of partial integration. Finally, we
get the following expression for the function $I(\tau)$:
\begin{equation}
  \label{ITfin}
  I(\tau)=\sum_{k=1}^{\infty}\frac{\tau^{k}}{k}
  \frac{(k!)^{2}}{(2k+1)!}-
  \sum_{k=1}^{\infty}\frac{k+2}{k}
  \frac{(k!)^{2}}{(2k+1)!}\tau^{k}
  \int_{0}^{1}dz\frac{z^{2k+1}}{D^{k}}.
\end{equation}
Note that the first term in Eq.~\eqref{ITfin} does not depend on
the charged lepton and neutrino mass parameters, $a$ and $b$. It
is this term that cancels the corresponding contribution of the
$\gamma-Z$ self-energy diagrams. The subsequent analysis of the
remaining contributions of the proper vertices diagrams can be
performed in the same manner as we have done it for the function
$I(\tau)$. Finally, we obtain that
\begin{equation*}
  f_{5}(q^{2})=f_{5}^{(\gamma-Z)}(q^{2})+
  f_{5}^{(\mathrm{prop. vert.})}(q^{2})=0
\end{equation*}
for any value of $q^{2}$ and for arbitrary charged lepton $a$ and
neutrino $b$ mass parameters. It should be noted that the
decomposition of the fermion electromagnetic vertex function in
terms of the four form factors was established previously with the
use of only general principles such as the Lorentz invariance and
hermicity of the electromagnetic current operator. We have
demonstrated the validity of this decomposition by means of the
direct calculations of the corresponding Feynman diagrams.

\section{Neutrino Electric Form Factor \label{NCF}}

In this section we study the massive neutrino electric form
factor. Using the results of the previous section for different
contributions to the neutrino vertex $\Lambda_{\mu}(q)$ we extract
in Eqs.~\eqref{L1}-\eqref{P14} the coefficients proportional to
$\gamma_{\mu}$-matrix that are, according to the
decomposition~\eqref{J}, the corresponding contributions to the
neutrino electric form factor $f_{Q}(q^2)$.

First of all we consider the contributions of the one-loop proper
vertices [Fig.~\ref{prverta}-\ref{prvertf}] to the neutrino
electric form factor. Using the well known mass shell identity
\begin{equation*}
  \bar{u}(p^{\prime})(p^{\prime}_{\mu}+p_{\mu})u(p)=
  \bar{u}(p^{\prime})(2m_{\nu}\gamma_{\mu}-
  i\sigma_{\mu\nu}q^{\nu})u(p),
\end{equation*}
and carrying out an integration over the virtual momenta within
the dimensional-regularization scheme (see for more details
Appendix~\ref{FeynInt}) we derive the exact expressions for the
contributions from the considered diagrams to the massive neutrino
electric form factor in terms of the definite integrals
\begin{equation*}
  f_Q^{(\mathrm{prop.vert.})}(q^2)=
  {\frac{eG_{F}}{4\pi^{2}\sqrt{2}}}M_{W}^{2}
  \sum_{i=1}^{6}
  \bar{f}_Q^{(i)}(q^2),
\end{equation*}
where
\begin{multline}
  \label{fQ1}
  \bar{f}^{(1)}_Q(q^2)=
  \omega\frac{\alpha}{2}+1+\frac{1-\alpha}{12}+
  \int_0^1 dz
  \int_0^z dy
  \ln \mathfrak{D}_1-
  \int_0^1 dz
  \int_0^z dy
  \big(
  a+b(1-z)^2+\tau(1-z+y(z-y))
  \big)
  \frac{1}{\mathfrak{D}_1}+
  \\
  \frac{1}{2}
  \int_0^1 dz
  \int_0^z dy
  \big(
  bz^2(a+b(1-z)^2)+a\tau y(z-y)+
  \\
  b\tau(2zy(z-y)(1-z)+5y(z-y)-z^2(1-z))+
  \tau^2y(z-y)(1-z+yz-y^2)
  \big)
  \left[
  \frac{1}{\mathfrak{D}_1(\alpha)}-\frac{1}{\mathfrak{D}_1}
  \right]-
  \\
  \frac{1}{2}
  \int_0^1 dz
  \int_0^z dy
  \left(
  a+b+6bz(1-z)+\tau(1-3z+6y(z-y))
  \right)
  \left[
  \ln \mathfrak{D}_1(\alpha)-\ln \mathfrak{D}_1
  \right],
\end{multline}
\begin{multline}
  \label{fQ2}
  \bar{f}^{(2)}_Q(q^2)=
  \frac{a+b}{2}
  \left(
  \frac{\omega}{2}+\frac{1}{2}+
  \int_0^1 dz
  \int_0^z dy
  \ln \mathfrak{D}_1(\alpha)
  \right)-
  \\
  \frac{1}{2}
  \int_0^1 dz
  \int_0^z dy
  \big(
  a^{2}+abz^{2}+b^{2}z^{2}-4abz+ab+
  (a+b)\tau y(z-y)
  \big)
  \frac{1}{\mathfrak{D}_1(\alpha)},
\end{multline}
\begin{equation}
  \label{fQ3}
  \bar{f}^{(3)}_Q(q^2)=
  \frac{a+b}{2}
  \left(
  -\frac{\omega}{2}-
  \int_0^1 dz
  \int_0^z dy
  \ln \mathfrak{D}_2(\alpha)
  \right)+
  b
  \int_0^1 dz
  \int_0^z dy
  \big(
  3az-az^{2}-2a+bz(1-z)
  \big)
  \frac{1}{\mathfrak{D}_2(\alpha)},
\end{equation}
\begin{multline}
  \label{fQ4}
  \bar{f}^{(4)}_Q(q^2)=
  -\omega\frac{3}{4}(1+\alpha)-1-
  3
  \int_0^1 dz
  \int_0^z dy
  \ln \mathfrak{D}_2+
  \int_0^1 dz
  \int_0^z dy
  \big(
  3bz(1-z)-\tau(z-y(z-y))
  \big)
  \frac{1}{\mathfrak{D}_2}-
  \\
  \frac{9}{2}
  \int_0^1 dz
  \int_0^z dy
  \left[
  (\mathfrak{D}_2(\alpha)+y(1-\alpha))\ln(\mathfrak{D}_2(\alpha)+y(1-\alpha))-
  \mathfrak{D}_2\ln \mathfrak{D}_2
  \right]-
  \\
  \int_0^1 dz
  \int_0^z dy
  \big(
  2b^2(1-z)^2(z(1-z)-y)-
  b\tau(y(z-y)(5z-3z^2-3y)+z(1-z)^2-y(2-y-y^2))-
  \\
  \tau^2y(z-y)(1-z+yz+y+y^2)
  \big)
  \left[
  \frac{1}{\mathfrak{D}_2(\alpha)+y(1-\alpha))}-\frac{1}{\mathfrak{D}_2}
  \right]+
  \\
  \frac{1}{2}
  \int_0^1 dz
  \int_0^z dy
  \big(
  3b(1-z^2)+\tau(4-6(z-y)+11y(z-y))
  \big)
  \left[
  \ln(\mathfrak{D}_2(\alpha)+y(1-\alpha))-\ln \mathfrak{D}_2
  \right]-
  \\
  \frac{b\tau}{2}
  \int_0^1 dz
  \int_0^z dy
  \big(
  bz(1-3z+z^2+z^3)-
  \tau y(z-y)(z+z^2-2y)
  \big)
  \left[
  \frac{1}{\mathfrak{D}_2}+\frac{1}{\mathfrak{D}_2(\alpha)}-
  \frac{2}{\mathfrak{D}_2(\alpha)+y(1-\alpha)}
  \right]+
  \\
  \frac{\tau}{4}
  \int_0^1 dz
  \int_0^z dy
  \big(
  b(9-13z+4z^2)-2\tau y(z-y)
  \big)
  \left[
  \ln \mathfrak{D}_2+\ln \mathfrak{D}_2(\alpha)-
  2\ln(\mathfrak{D}_2(\alpha)+y(1-\alpha))
  \right]+
  \\
  \frac{3\tau}{4}
  \int_0^1 dz
  \int_0^z dy
  \left[
  \mathfrak{D}_2\ln \mathfrak{D}_2+\mathfrak{D}_2(\alpha)\ln \mathfrak{D}_2(\alpha)-
  2(\mathfrak{D}_{2}(\alpha)+y(1-\alpha))
  \ln(\mathfrak{D}_{2}(\alpha)+y(1-\alpha))
  \right],
\end{multline}
\begin{multline}
  \label{fQ56}
  \bar{f}^{(5)+(6)}_Q(q^2)=
  \int_0^1 dz
  \int_0^z dy
  (a-bz)
  \frac{1}{\mathfrak{D}_2(\alpha)+y(1-\alpha)}-
  \\
  b
  \int_0^1 dz
  \int_0^z dy
  (1-z)
  \big(
  (1-z)(a-bz)-\tau y(z-y)
  \big)
  \left[
  \frac{1}{\mathfrak{D}_2(\alpha)+y(1-\alpha)}-\frac{1}{\mathfrak{D}_2(\alpha)}
  \right]-
  \\
  \frac{1}{2}
  \int_0^1 dz
  \int_0^z dy
  (a+5b-6bz)
  \left[
  \ln(\mathfrak{D}_2(\alpha)+y(1-\alpha))-\ln \mathfrak{D}_2(\alpha)
  \right].
\end{multline}
Here
\begin{align*}
  \mathfrak{D}_{1}(\alpha) &=
  \alpha+(a-\alpha)z-bz(1-z)+\tau y(z-y),
  &
  \mathfrak{D}_{1} &=
  \mathfrak{D}_{1}(\alpha=1)=1+(a-1)z-bz(1-z)+\tau y(z-y),
  \\
  \mathfrak{D}_{2}(\alpha) &=
  a+(\alpha-a)z-bz(1-z)+\tau y(z-y),
  &
  \mathfrak{D}_{2} &=
  \mathfrak{D}_{2}(\alpha=1)=a+(1-a)z-bz(1-z)+\tau y(z-y).
\end{align*}
Note that the values of the mass parameters of the charged lepton
($a$) and neutrino ($b$) are taken into account explicitly in
Eqs.~\eqref{fQ1}-\eqref{fQ56}. The gauge parameter $\alpha$ and
$q^2$ are arbitrary in these formulae.

The contributions of the $\gamma-Z$ self-energy diagrams
(Fig.~\ref{gZverta}-\ref{gZverth}) to the electric form factor can
be obtained using Eqs.~\eqref{L7-14} and \eqref{P7-14}. Thus, one
obtains
\begin{equation}
  \label{fQ7-14}
  f_Q^{(j)}(q^2)=
  \frac{g}{4\cos\theta_{W}}
  \frac{A^{(j)}(\alpha,q^2)+B^{(j)}(\alpha,q^2)}{q^2-M_{Z}^{2}},
  \quad
  j=7,\dots,14.
\end{equation}
Using explicit form of the functions $A^{(j)}(\alpha,q^2)$
[Eqs.~\eqref{A7}-\eqref{A14}] and $B^{(j)}(\alpha,q^2)$
[Eqs.~\eqref{B7}-\eqref{B14}] as well as Eq.~\eqref{fQ7-14}, one
can also derive the expressions for the contributions of the
$\gamma-Z$ self-energy diagrams at arbitrary values of the gauge
parameter $\alpha$ and $q^2\not=0$.

\subsection{Neutrino Electric Charge in
Arbitrary Gauge \label{NEC}}

In this section we consider the neutrino electric charge. At zero
momentum transfer the sum of the contributions to the electric
form factor determines the neutrino charge, $f_{Q}(0)=Q$. Our goal
is to find its total value for the massive neutrino
\begin{equation*}
  Q=\sum_{i=1}^{6} Q^{(i)}(a,b,\alpha)+\sum_{j=7}^{14}
  Q^{(j)}(a,b,\alpha)
\end{equation*}
and to study the mass ($a$ and $b$) and gauge-fixing ($\alpha$ and
$\alpha_{Z}$) parameters dependence of the contributions from the
different Feynman diagrams depicted in
Figs.~\ref{prverta}-\ref{prvertf} and \ref{gZverta}-\ref{gZverth}.

First we consider the one-loop contributions to the neutrino
charge which arise from the proper vertices diagrams in
Fig.~\ref{prverta}-\ref{prvertf}. Using of the more general
formulae for the electric neutrino form factor
Eqs.~\eqref{fQ1}-\eqref{fQ56}, we obtain the exact expressions for
the contributions from the considered diagrams to the massive
neutrino charge in terms of the definite integrals:
\begin{multline}
  \label{Q1}
  Q^{(1)}(a,b,\alpha)=
  {\frac{eG_{F}}{4\pi^{2}\sqrt{2}}}M_{W}^{2} \bigg\{
  \omega{\frac{\alpha}{2}}+1+{\frac{1-\alpha}{12}}+ \int_{0}^{1} dz
  (1-z) \ln D-
  \\
  \int_{0}^{1} dz (1-z)
  (a+bz^{2})
  {\frac{1}{D}}+
  {\frac{b}{2}}\int_{0}^{1} dz
  (1-z)^{3}(a+bz^{2})
  \left[
  {\frac{1}{D}_{\alpha}}-{\frac{1}{D}}
  \right]-
  \\
  {\frac{1}{2}}\int_{0}^{1} dz (1-z) (a+b+6bz(1-z)) \left[ \ln
  D_{\alpha}-\ln D \right]+3\int_{0}^{1} dz (1-z)
  \left[
  D_{\alpha}\ln D_{\alpha}-D\ln D
  \right]
  \bigg\},
\end{multline}
\begin{multline}
  \label{Q2}
  Q^{(2)}(a,b,\alpha)=
  {\frac{eG_{F}}{4\pi^{2}\sqrt{2}}}M_{W}^{2} \bigg\{ {\frac{a+b}{2}}
  \Big(
  {\frac{\omega}{2}}+{\frac{1}{2}}+
  \int_{0}^{1} dz (1-z)
  \ln D_{\alpha}
  \Big)-
  \\
  {\frac{1}{2}}\int_{0}^{1} dz (1-z)
  (a^{2}+abz^{2}+b^{2}z^{2}-4abz+ab)
  {\frac{1}{D_{\alpha}}}
  \bigg\},
\end{multline}
\begin{equation}
  \label{Q3}
  Q^{(3)}(a,b,\alpha)=
  {\frac{eG_{F}}{4\pi^{2}\sqrt{2}}}M_{W}^{2}
  \bigg\{
  {\frac{a+b}{2}}
  \Big(
  -{\frac{\omega}{2}}
  -\int_{0}^{1} dz\thinspace z
  \ln D_{\alpha}
  \Big)+
  b\int_{0}^{1} dz {\kern 3 pt} z {\kern 3 pt}
  (3az-az^{2}-2a+bz(1-z))
  {\frac{1}{D_{\alpha}}}
  \bigg\},
\end{equation}
\begin{multline}
  \label{Q4}
  Q^{(4)}(a,b,\alpha)=
  {\frac{eG_{F}}{4\pi^{2}\sqrt{2}}}M_{W}^{2} \bigg\{
  -\omega{\frac{3}{4}}(1+\alpha)-1-3\int_{0}^{1} dz\thinspace z
  \ln D + 3b\int_{0}^{1} dz\thinspace z^{2}(1-z) {\frac{1}{D}}-
  \\
  {\frac{9}{2}}\int_{0}^{1} dz \int_{0}^{z} dy
  \left[
  (D_{\alpha}+y(1-\alpha))\ln (D_{\alpha}+y(1-\alpha))-
  D\ln D
  \right]-
  \\
  2b^{2}\int_{0}^{1} dz \int_{0}^{z} dy
  (1-z)^{2}(z(1-z)-y)
  \left[
  {\frac{1}{D_{\alpha}+y(1-\alpha))}}-{\frac{1}{D}}
  \right]+
  \\
  {\frac{3}{2}}b
  \int_{0}^{1} dz \int_{0}^{z} dy
  (1-z^{2})
  \left[
  \ln (D_{\alpha}+y(1-\alpha))-\ln D
  \right]
  \bigg\},
\end{multline}
\begin{multline}
  \label{Q56}
  Q^{(5)+(6)}(a,b,\alpha)=
  {\frac{eG_{F}}{4\pi^{2}\sqrt{2}}}M_{W}^{2} \bigg\{ \int_{0}^{1} dz
  \int_{0}^{z} dy (a-bz) {\frac{1}{D_{\alpha}+y(1-\alpha)}}-
  \\
  b\int_{0}^{1} dz \int_{0}^{z} dy
  (1-z)^{2}(a-bz)
  \left[
  {\frac{1}{D_{\alpha}+y(1-\alpha)}}-{\frac{1}{D_{\alpha}}}
  \right]-
  \\
  {\frac{1}{2}}\int_{0}^{1} dz \int_{0}^{z} dy
  (a+5b-6bz)
  \left[
  \ln (D_{\alpha}+y(1-\alpha))-\ln D_{\alpha}
  \right]
  \bigg\}.
\end{multline}

The integral expressions of Eqs.~\eqref{Q1}-\eqref{Q56} for
different proper vertices contributions to the neutrino charge
exactly account for the charged lepton and neutrino mass
parameters, $a$ and $b$, and also for the gauge-fixing parameter
$\alpha$. We have calculated the integrals in
Eqs.~\eqref{Q1}-\eqref{Q56}, however the results, being expressed
in elementary functions, are rather cumbersome. Therefore, we also
perform corresponding integrations in the first two terms of
expansion over the neutrino mass parameter $b$ for arbitrary
values of the charged lepton mass parameter $a$ and the
gauge-fixing parameter $\alpha$. In this case the sum of the
proper vertices diagrams to the neutrino electric charge can be
written as
\begin{equation}
  \label{q1-6}
  Q^{(\mathrm{prop. vert.})}(a,b,\alpha)=
  {\frac{eG_{F}}{4\pi^{2}\sqrt{2}}}M_{W}^{2}
  \sum_{i=1}^{6}\{q^{(i)}_{0}(a,\alpha)+bq^{(i)}_{1}(a,\alpha)+
  \mathcal{O}(b^{2})\}.
\end{equation}
For $q^{(i)}_{0}(a,\alpha)$ we obtain
\begin{multline}
  \label{Q01}
  q^{(1)}_{0}(a,\alpha)=
  \omega{\frac{\alpha}{2}}-
  {\frac{1}{4(1-a)^{2}(\alpha-a)^{2}}}
  \big(
  -3{a}^{2}+4{a}^{2}\alpha\ln a-5{a}^{2}\alpha+ 2{\alpha}^{3}
  +3{a}^{3}- 3{a}^{3}\ln a-2{\alpha}^{3}\ln \alpha+
  \\
  {\alpha}^{2}a\ln \alpha+
  2{a}^{2}{\alpha}^{3}-
  3{a}^{3}{\alpha}^{2}+
  \alpha{a}^{4}+
  6{a}^{2}{\alpha}^{2}+6a\alpha-3{\alpha}^{2}- 4{\alpha}^{3}a-
  2{a}^{3}\alpha +4{\alpha}^{3}a\ln \alpha- 2{a}^{4}\alpha\ln a-
  \\
  2{\alpha}^{3}{a}^{2}\ln \alpha+
  4\alpha{a}^{3}\ln a-
  6{a}^{2}{\alpha}^{2}\ln a+ {\alpha}^{2}{a}^{3}\ln \alpha-
  2{\alpha}^{2}{a}^{2}\ln \alpha+ 3{a}^{3}{\alpha}^{2}\ln a
  \big),
\end{multline}
\begin{equation}
  \label{Q02}
  q^{(2)}_{0}(a,\alpha)=
  \omega{\frac{a}{4}}+ {\frac{a}{8(\alpha-a)^2}}
  \big(
  2{\alpha}^{2}\ln \alpha + 4a\alpha-
  4a\alpha\ln \alpha - {\alpha}^{2}+2{a}^{2}\ln a-3{a}^{2}
  \big),
\end{equation}
\begin{equation}
  \label{Q03}
  q^{(3)}_{0}(a,\alpha)=
  -\omega{\frac{a}{4}}- {\frac{a}{8(\alpha-a)^2}}
  \big(
  2{\alpha}^{2}\ln \alpha + 4a\alpha-
  4a\alpha\ln \alpha - {\alpha}^{2}+2{a}^{2}\ln a-3{a}^{2}
  \big),
\end{equation}
\begin{multline}
  \label{Q04}
  q^{(4)}_{0}(a,\alpha)=
  -\omega{\frac{3}{4}}(1+\alpha)-
  {\frac{1}{8(1-a)^{2}(\alpha-a)(1-\alpha)}}
  \big(
  {\alpha}^{2}a-4{a}^{2}- 6{\alpha}^{3}\ln \alpha-6{a}^{3}\ln a-
  \\
  11{a}^{2}\alpha+5{\alpha}^{3}+5{a}^{3}+
  5{a}^{2}{\alpha}^{3}-5{a}^{3}{\alpha}^{2}+ 10{a}^{2}{\alpha}^{2}+
  10a\alpha-6{\alpha}^{2}- 10{\alpha}^{3}a-
  \\
  6{\alpha}^{3}{a}^{2}\ln\alpha- 12{a}^{2}{\alpha}^{2}\ln a+
  12{a}^{2}\alpha\ln a+ 12{\alpha}^{3}a\ln \alpha+
  6{a}^{3}{\alpha}^{2}\ln a+\alpha-a
  \big),
\end{multline}
\begin{multline}
  \label{Q056}
  q^{(5)+(6)}_{0}(a,\alpha)=
  {\frac{a}{4(\alpha-a)^{2}(1-a)(1-\alpha)}}
  \big(
  2a\alpha\ln\alpha+
  {a}^{2}\alpha- {\alpha}^{3}+4a\alpha\ln a-
  {a}^{2}{\alpha}^{2}-a\alpha+{\alpha}^{2}+{\alpha}^{3}a-
  \\
  3{\alpha}^{2}\ln \alpha+ {a}^{2}{\alpha}^{2}\ln a+
  2{a}^{2}\alpha\ln a+ 4{\alpha}^{2}a\ln\alpha-
  {\alpha}^{2}{a}^{2}\ln \alpha-4a{\alpha}^{2}\ln a-
  2{a}^{2}\alpha\ln \alpha-3{a}^{2}\ln a
  \big).
\end{multline}
Each of the coefficients $q^{(i)}_{0}(a,\alpha)$, if considered
separately, depends on the gauge-fixing parameter $\alpha$ and all
of them [except for $q^{(5)}_{0}(a,\alpha)$ and
$q^{(6)}_{0}(a,\alpha)$] are divergent. Note that, according to
the expansion given by Eq.~\eqref{q1-6}, the sum
\[
  \sum_{i=1}^{6}q^{(i)}_{0}(a,\alpha)
\]
determines the proper vertex contribution to the charge in the
limit of the massless neutrino.

It should also be noted that the two diagrams of
Fig.~\ref{prverte} and \ref{prvertf} are convergent for every
value of the gauge-fixing parameter $\alpha$. One can check this
statement using Eq.~\eqref{L56}. Indeed, these diagrams have the
superficial degree of divergence equal to $-1$ and hence converge
\cite{Wei96p500}. Therefore the according contributions to the
electric charge must be finite as it is also shown by
Eqs.~\eqref{Q56} and \eqref{Q056}. Here we find a discrepancy with
the corresponding results of Ref.~\cite{CabBerVidZep00} where the
massless neutrino charge is calculated and the contributions of
these two diagrams contain the ultraviolet divergencies.

The next order over the neutrino mass parameter $b$ of the proper
vertex diagrams' contributions to the neutrino charge can be
obtained if one expands the integrands in
Eqs.~\eqref{Q1}-\eqref{Q56}, keeps the terms proportional to $b$,
and then carries out the integration. Taking into account that the
functions $D$ and $D_{\alpha}$ also depend on $b$, we find that
\begin{equation}
  \label{sum_q_i}
  \sum_{i=1}^{6}q^{(i)}_{1}(a,\alpha)=0.
\end{equation}
Thus, due to the fact that, as it is shown below in this section,
the $\gamma - Z$ self-energy contribution does not depend on the
neutrino mass, it follows that the neutrino charge term
proportional to the neutrino mass parameter $b$ is zero.

Now let us turn to the $\gamma-Z$ self-energy contributions to the
neutrino electric charge. The corresponding Feynman diagrams are
depicted in Fig.~\ref{gZverta}-\ref{gZverth}. Using
Eq.~\eqref{P7-14}, which presents the decomposition of the
functions $\Pi_{\mu\nu}^{(j)}(q)$, as well as the explicit form of
the functions $A^{(j)}(\alpha,q^2)$ [Eqs.~\eqref{A7}-\eqref{A14}],
we find that
\begin{equation*}
  A^{(j)}(\alpha,q^{2}=0)=0.
\end{equation*}
Therefore only the terms proportional to $B^{(j)}(\alpha,0)$ are
responsible for the neutrino electric charge in the $\gamma-Z$
self-energy Feynman diagrams and we have
\begin{equation*}
  Q^{(\gamma-Z)}(\alpha)=\sum_{j=7}^{14}Q^{(j)}(\alpha)=
  -{\frac{g}{4M_{Z}^{2}\cos\theta_{W}}}
  \sum_{j=7}^{14}B^{(j)}(\alpha,0).
\end{equation*}
For each of the contributions $Q^{(j)}(\alpha)$ ($j=7,\dots,14$)
from Eq.~\eqref{fQ7-14} we obtain
\begin{equation}
  \label{Q7}
  Q^{(7)}(\alpha)=
  {\frac{eG_{F}}{4\pi^{2}\sqrt{2}}}M_{W}^{2}\cos^{2}\theta_{W}
  \bigg\{
  \omega \left[ 3+{\frac{3}{4}}\alpha(1+\alpha)
  \right]-
  1-{\frac{5\alpha}{8}}- {\frac{5\alpha^{2}}{8}}-
  {\frac{3\alpha^{3}}{4}} {\frac{\ln \alpha}{(1-\alpha)}}
  \bigg\},
\end{equation}
\begin{equation}
  \label{Q8}
  Q^{(8)}(\alpha)=
  {\frac{eG_{F}}{4\pi^{2}\sqrt{2}}}M_{W}^{2}\sin^{2}\theta_{W}
  \left\{
  \omega{\frac{3+\alpha}{4}}-
  {\frac{5+\alpha}{8}}- {\frac{\alpha}{2}}
  \left(
  1+{\frac{\alpha}{2}}
  \right)
  {\frac{\ln \alpha}{(1-\alpha)}}
  \right\},
\end{equation}
\begin{equation}
  \label{Q9}
  Q^{(9)}(\alpha)={\frac{eG_{F}}{4\pi^{2}\sqrt{2}}}M_{W}^{2}
  (\cos^{2}\theta_{W}-\sin^{2}\theta_{W})
  {\frac{1}{2}}
  \left\{
  -\omega\alpha+\alpha-\alpha\ln \alpha
  \right\},
\end{equation}
\begin{equation}
  \label{Q10}
  Q^{(10)}(\alpha)=
  {\frac{eG_{F}}
  {4\pi^{2}\sqrt{2}}}M_{W}^{2}\cos^{2}\theta_{W}
  \left\{
  -{\frac{3}{4}}\omega(3+\alpha^{2})+
  {\frac{3}{8}}+ {\frac{5\alpha^{2}}{8}}-
  {\frac{3}{4}}\alpha^{2}\ln \alpha
  \right\},
\end{equation}
\begin{equation}
  \label{Q1112}
  Q^{(11)+(12)}(\alpha)={\frac{eG_{F}}{4\pi^{2}\sqrt{2}}}M_{W}^{2}
  \cos^{2}\theta_{W} {\frac{1}{2}} \left\{
  -\omega\alpha+\alpha-\alpha\ln \alpha \right\},
\end{equation}
\begin{equation}
  \label{Q13}
  Q^{(13)}(\alpha)={\frac{eG_{F}}{4\pi^{2}\sqrt{2}}}M_{W}^{2}
  (\sin^{2}\theta_{W}-\cos^{2}\theta_{W})
  {\frac{1}{2}}
  \left\{
  -\omega\alpha+\alpha-\alpha\ln \alpha
  \right\},
\end{equation}
\begin{equation}
  \label{Q14}
  Q^{(14)}(\alpha)=0.
\end{equation}
It is worth to mention that each of the contributions
$Q^{(j)}(\alpha)$ turns out to be independent on the neutrino
($m_{\nu}$) and charged lepton ($m_{\ell}$) masses. There is also
no dependence on the masses $m_{f}$ of the virtual fermions that
circulate in the $\gamma-Z$ self-energy diagrams because of the
properties of the $\gamma$-matrix algebra specified in
Appendix~\ref{FeynInt}. The dependence on the gauge-fixing
parameter $\alpha_{Z}$ also cancels out within each of the
contributions. Note that prior the integrations in
Eqs.~\eqref{P7}-\eqref{P14} is carried out the $\alpha_{Z}$
dependence drops out of each of the electric form factor
contributions to the vertex function at arbitrary momentum
transfer $q^2$.

Finally, for the sum of all $\gamma-Z$ self-energy contributions
to the neutrino electric charge we have
\begin{equation}
  \label{Q7-14}
  Q^{(\gamma-Z)}(\alpha)=
  {\frac{eG_{F}}{4\pi^{2}\sqrt{2}}}M_{W}^{2}
  \left\{
  {\frac{3+\alpha}{4}}\omega-{\frac{5+\alpha}{8}}-
  {\frac{\alpha\ln \alpha}{2(1-\alpha)}} \left( 1+{\frac{\alpha}{2}}
  \right)
  \right\}.
\end{equation}
Some remarks should be made with respect to the divergent parts in
Eqs.~\eqref{Q01}-\eqref{Q04} and \eqref{Q7-14}. The sum of all the
coefficients in $\omega$ terms is zero, i.e. the electric charge
of a massive neutrino vanishes for every number of dimensions $N$.
The same property of the electric charge of a massless neutrino
was determined in Ref.~\cite{LucRosZep84}.

Now we can complete investigation of the neutrino charge in the
zeroth order of the expansion over the neutrino mass parameter $b$
summing together the contributions from the proper vertices given
by Eqs.~\eqref{q1-6}-\eqref{Q056},
\begin{equation}
  \label{p_v_0}
  Q^{(\mathrm{prop.vert.})}(a,0,\alpha)=
  {\frac{eG_{F}}{4\pi^{2}\sqrt{2}}}M_{W}^{2}
  \sum_{i=1}^{6}
  q^{(i)}_{0}(a,\alpha),
\end{equation}
and $Q^{(\gamma-Z)}(\alpha)$ given by Eq.~\eqref{Q7-14}.

As a result we obtain that the neutrino electric charge in the
zeroth order in the neutrino mass vanishes for every gauge in
agreement with the final results of
Refs.~\cite{CabBerVidZep00,LucRosZep84}, where the calculations of
the neutrino electric charge have been performed in the limit of
vanishing neutrino mass.

\subsection{Neutrino Electric Charge in the
't~Hooft-Feynman Gauge}

Within the 't~Hooft-Feynman gauge it is possible to show
explicitly that at the one-loop level the neutrino electric charge
is zero for arbitrary mass of neutrino. The gauge-fixing parameter
$\alpha=1$ in this gauge. Summing up the contributions of all
relevant diagrams [Eqs.~\eqref{Q1}-\eqref{Q56} and \eqref{Q7-14}]
we receive the exact expression for the neutrino charge at
arbitrary values of the charged lepton and neutrino mass
parameters, $a$ and $b$, while $\alpha=1$:
\begin{multline}
  \label{QtHFg}
  Q(a,b,\alpha=1)={\frac{eG_{F}}{4\pi^{2}\sqrt{2}}}M_{W}^{2}
  \bigg\{
  {\frac{1}{2}}\int_{0}^{1} dz
  [a(-(a+2)+z(a+4))+
  \\
  ab(-1+z+z^{2}-z^{3})+
  2bz^{2}(1-2z)+b^{2}z^{2}(1-z)]
  {\frac{1}{D}}+
  \\
  \int_{0}^{1} dz (1-4z)\ln D+
  {\frac{a+b}{2}}
  \left(
  {\frac{1}{2}}+
  \int_{0}^{1} dz (1-2z)\ln D
  \right)
  \bigg\}.
\end{multline}
To analyze Eq.~\eqref{QtHFg} we use the formulae
\begin{equation}
  \label{int_12}
  \int_{0}^{1} dz \ln D = -1+ \int_{0}^{1} dz
  (a-bz^{2}){\frac{1}{D}},
  \quad
  \int_{0}^{1} dz\thinspace z \ln D = -{\frac{1}{4}}+
  {\frac{1}{2}}\int_{0}^{1} dz\thinspace z(a-bz^{2}){\frac{1}{D}},
\end{equation}
that can be proven by means of partial integration. Substituting
Eq.~\eqref{int_12} into Eq.~\eqref{QtHFg} we obtain that the
neutrino electric charge vanishes for arbitrary neutrino mass in
the considered gauge.

\subsection{Neutrino Charge Radius \label{NCR}}

Using the closed expressions for the contributions to the neutrino
electric form factor obtained above in this Section, it is
possible also to derive the neutrino charge radius. Accounting for
the next-to-leading term in the  $q^2$-expansion of the
contributions in Eqs.~\eqref{fQ1}-\eqref{fQ56} and \eqref{fQ7-14},
\begin{equation*}
  f_{Q}(q^2)=f_{Q}(0)+q^2\frac{df_Q}{dq^2}(0)+\dotsb,
\end{equation*}
one can obtain the value of the massive neutrino charge radius as
\begin{equation*}
  \langle
  r_\nu^2
  \rangle=
  -6\frac{df_Q}{dq^2}(0)
\end{equation*}
and study its dependencies on the gauge and mass parameters. Here
we should like to note that the problem of the massless neutrino
charge radius has been discussed in details in
\cite{LucRosZep85,BerCabPapVid00}.

\section{Neutrino Magnetic Moment \label{NMM}}

According to the general decomposition of the neutrino
electromagnetic vertex function $\Lambda_{\mu}(q)$ given by
Eq.~\eqref{J}, the neutrino dipole magnetic form factor
$f_{M}(q^2)$ is the coefficient in the term proportional to
$i\sigma_{\mu\nu}q^{\nu}$. In this section we first determine
$f_{M}(q^2)$ and then calculate at $q^{2}=0$ the neutrino magnetic
moment accounting for the two mass parameters ($a$ and $b$) and
for the gauge-fixing parameter ($\alpha$) as well,
\begin{equation}
  \label{mu}
  \mu(a,b,\alpha)=f_{M}(q^{2}=0).
\end{equation}
Note that the Feynman diagrams in Fig.~\ref{gZverta}-\ref{gZverth}
do not contribute to the neutrino magnetic moment. Thus, the total
one-loop value for the neutrino magnetic moment is given by
\begin{equation}
  \label{mu_1-6}
  \mu(a,b,\alpha)=\sum_{i=1}^{6}\mu^{(i)}(a,b,\alpha),
\end{equation}
where $\mu^{(i)}(a,b,\alpha)$ are the contributions to the
magnetic moment from the corresponding diagrams shown in
Fig.~\ref{prverta}-\ref{prvertf}.

We treat the neutrino magnetic moment in the similar way as we
have analyzed the neutrino electric charge. Using
Eqs.~\eqref{L1}-\eqref{L56} for each of the contributions to the
neutrino magnetic moment we receive
\begin{multline}
  \label{mu1}
  \mu^{(1)}(a,b,\alpha)=
  {\frac{eG_{F}}{4\pi^{2}\sqrt{2}}}m_{\nu}
  \bigg\{
  \int_{0}^{1} dz
  z(1-z^{2}){\frac{1}{D}}-
  {\frac{1}{2}}\int_{0}^{1} dz
  (1-z)^{3}(a-bz)
  \left[
  {\frac{1}{D}_{\alpha}}-{\frac{1}{D}}
  \right]-
  \\
  {\frac{1}{2}}\int_{0}^{1} dz
  (1-z)(1-3z)
  \left[
  \ln D_{\alpha}-\ln D
  \right]
  \bigg\},
\end{multline}
\begin{equation}
  \label{mu2}
  \mu^{(2)}(a,b,\alpha)=
  {\frac{eG_{F}}{4\pi^{2}\sqrt{2}}}m_{\nu}
  {\frac{1}{2}}\int_{0}^{1} dz (1-z)
  (-3az+az^{2}+2a-bz(1-z))
  {\frac{1}{D_{\alpha}}},
\end{equation}
\begin{equation}
  \label{mu3}
  \mu^{(3)}(a,b,\alpha)=
  {\frac{eG_{F}}{4\pi^{2}\sqrt{2}}}m_{\nu}
  {\frac{1}{2}}\int_{0}^{1}dz\thinspace z
  (-3az+az^{2}+2a-bz(1-z)) {\frac{1}{D_{\alpha}}},
\end{equation}
\begin{multline}
  \label{mu4}
  \mu^{(4)}(a,b,\alpha)=
  {\frac{eG_{F}}{4\pi^{2}\sqrt{2}}}m_{\nu}
  \bigg\{
  {\frac{1}{2}}\int_{0}^{1} dz\thinspace z^{2}(1+2z)
  {\frac{1}{D}}+
  \\
  {\frac{b}{2}}\int_{0}^{1} dz \int_{0}^{z} dy
  (1-z)^{2}(z(1-z)-2y)
  \left[
  {\frac{1}{D_{\alpha}+y(1-\alpha)}}-{\frac{1}{D}}
  \right]+
  \\
  {\frac{1}{2}}
  \int_{0}^{1} dz \int_{0}^{z} dy
  (-2+9z-4z^{2}-6y)
  \left[
  \ln (D_{\alpha}+y(1-\alpha))-\ln D
  \right]
  \bigg\},
\end{multline}
\begin{multline}
  \label{mu56}
  \mu^{(5)+(6)}(a,b,\alpha)=
  {\frac{eG_{F}}{4\pi^{2}\sqrt{2}}}m_{\nu}
  \bigg\{
  \int_{0}^{1} dz
  \int_{0}^{z} dy\thinspace y {\frac{1}{D_{\alpha}+y(1-\alpha)}}+
  \\
  {\frac{1}{2}}\int_{0}^{1} dz \int_{0}^{z} dy
  (1-z)^{2}(a-bz)
  \left[
  {\frac{1}{D_{\alpha}+y(1-\alpha)}}-{\frac{1}{D_{\alpha}}}
  \right]+
  \\
  {\frac{1}{2}}\int_{0}^{1} dz \int_{0}^{z} dy
  (2-3z)
  \left[
  \ln (D_{\alpha}+y(1-\alpha))-\ln D_{\alpha}
  \right]
  \bigg\}.
\end{multline}
It should be noted that these formulae exactly account for
dependencies on the neutrino and charged lepton mass parameters
($a$ and $b$) and the gauge-fixing parameter ($\alpha$).

To proceed further with the analytical calculations we expand the
contributions to the neutrino magnetic moment
[Eqs.~\eqref{mu1}-\eqref{mu56}] over the neutrino mass parameter
$b$ and consider the first two terms. Then from Eq.~\eqref{mu_1-6}
we obtain
\begin{equation}
  \label{barmu}
  \mu(a,b,\alpha)={\frac{eG_{F}}{4\pi^{2}\sqrt{2}}}m_{\nu}
  \sum_{i=1}^{6}\{\bar\mu_{0}^{(i)}(a,\alpha)+
  b\bar\mu_{1}^{(i)}(a,\alpha)+\mathcal{O}(b^{2})\}.
\end{equation}
For each of the coefficients $\bar{\mu}_{0}^{(i)}(a,\alpha)$ we
have found the exact expressions in terms of algebraic functions,
however they are again rather cumbersome. Therefore, let us
consider more compact expressions for
$\bar\mu_{0}^{(i)}(a,\alpha)$ that can be obtained in expansion
over the charged lepton mass parameter $a$. Thus, accounting for
the terms up to the second order in $a$ we derive for the
coefficients $\bar{\mu}_{0}^{(i)}(a,\alpha)$
\begin{equation}
  \label{mu01}
  \bar{\mu}_{0}^{(1)}(a,\alpha)=
  {\frac{2}{3}}+
  {\frac{10-3\alpha +6\ln a -6\ln \alpha}
  {6\alpha}}a+\mathcal{O}(a^{2}),
\end{equation}
\begin{equation}
  \label{mu02}
  \bar{\mu}_{0}^{(2)}(a,\alpha)=
  -{\frac{5+3\ln a-3\ln \alpha }{3\alpha}}a +\mathcal{O}(a^{2}),
\end{equation}
\begin{equation}
  \label{mu03}
  \bar{\mu}_{0}^{(3)}(a,\alpha)=
  {\frac{5a}{12\alpha}} +\mathcal{O}(a^{2}),
\end{equation}
\begin{equation}
  \label{mu04}
  \bar{\mu}_{0}^{(4)}(a,\alpha)=
  {\frac{2-7\alpha-3\alpha\ln \alpha+5\alpha^{2}}
  {6(1-\alpha)^{2}}}-
  {\frac{9-12\alpha+\ln \alpha+5\alpha\ln \alpha+3\alpha^2}
  {12(1-\alpha)^{2}}}a +\mathcal{O}(a^{2}),
\end{equation}
\begin{equation}
  \label{mu056}
  \bar{\mu}_{0}^{(5)+(6)}(a,\alpha)=
  {\frac{1-\alpha+\alpha\ln \alpha}{2(1-\alpha)^{2}}}-
  {\frac{5-16\alpha-\alpha\ln \alpha+
  11\alpha^{2}-5\alpha^{2}\ln\alpha}
  {12\alpha (1-\alpha)^{2}}}a+
  \mathcal{O}(a^{2}).
\end{equation}
Eqs.~\eqref{mu01}-\eqref{mu056} together with Eq.~\eqref{barmu}
yield a value of the magnetic moment in the limit $b\to 0$ that
corresponds to the case of a light neutrino. We may compare our
$\bar{\mu}_{0}^{(i)}(a,\alpha)$ calculations with the results of
Ref.~\cite{CabBerVidZep00}. Our results for the contributions
$\mu_{0}^{(4)}(a,\alpha)$, $\mu_{0}^{(5)}(a,\alpha)$, and
$\mu_{0}^{(6)}(a,\alpha)$ disagree with those of the above cited
paper. The Feynman diagrams corresponding to the contributions
$i=5$, $6$ contain the unphysical charged scalar boson. This boson
contributions should disappear in the unitary gauge when the gauge
parameter $\alpha=\infty$. Thus, the contributions to the magnetic
moment from these two diagrams must vanish in the limit
$\alpha\rightarrow\infty$. This is exactly what we get from
Eq.~\eqref{mu056}. However, the similar expression from
Ref.~\cite{CabBerVidZep00} does not depend on the gauge parameter
at all. An argument in favor of our results can be also obtained
if one considers the value of the neutrino magnetic moment within
the unitary gauge. Indeed, it is easy to show that using the
results of Ref.~\cite{CabBerVidZep00} it is not possible to get
the right value for the neutrino magnetic moment within this
gauge. In the unitary gauge, only the diagrams shown in
Fig.~\ref{prverta} and \ref{prvertd} contribute to the neutrino
magnetic moment. The results for these two contributions, that can
be obtained using the corresponding formulae presented in
\cite{CabBerVidZep00}, are
\begin{equation}
  \label{mu1wrong}
  \mu_{0}^{(1)}=
  {\frac{eG_{F}}{4\pi^{2}\sqrt{2}}}m_{\nu}
  \left\{
  {\frac{2}{3}}-{\frac{a}{2}}+\mathcal{O}(a^{2})
  \right\},
\end{equation}
\begin{equation}
  \label{mu4wrong}
  \mu_{0}^{(4)}=
  {\frac{eG_{F}}{4\pi^{2}\sqrt{2}}}m_{\nu}
  \left\{
  {\frac{7}{12}}+\mathcal{O}(a^{2})
  \right\}.
\end{equation}
The sum of the leading terms in Eqs.~\eqref{mu1wrong} and
\eqref{mu4wrong} differs from the well known result for the
neutrino magnetic moment calculation (see, for example,
Ref.~\cite{LeeShr77}):
\begin{equation*}
  \mu=\frac{3eG_{F}m_{\nu}}{8\pi^{2}\sqrt{2}}.
\end{equation*}
This fact points out that the contributions of the three diagrams
shown in Fig.~\ref{prvertd}-\ref{prvertf} are calculated in
\cite{CabBerVidZep00} with incorrect gauge-fixing parameter
$\alpha$ dependence. Our calculation shows that to the leading
order in $m_{\nu}$ each of these three contributions of the
diagrams in Fig.~\ref{prvertd}-\ref{prvertf} are gauge-fixing
parameter dependent. Note that calculations performed in
\cite{CabBerVidZep00} provide the correct results only within the
't~Hooft-Feynman gauge.

Let us now consider the value of the neutrino magnetic moment in
the "zeroth" order in the expansion over the neutrino mass
parameter $b$ taking into account all the contributions. The sum
of the coefficients \eqref{mu01}-\eqref{mu056} is found to be
independent on the gauge parameter $\alpha$. The straightforward
calculation of the neutrino magnetic moment in the limit $b\to 0$
yields
\begin{equation}
  \label{mu0tot}
  \mu_{0}(a,\alpha)=
  {\frac{eG_{F}}{4\pi^{2}\sqrt{2}}}m_{\nu}
  {\frac{3}{4(1-a)^{3}}}
  (2-7a+6a^{2}-2a^{2}\ln a-a^{3})
\end{equation}
that is in agreement with Ref.~\cite{CabBerVidZep00}.

Considering the next-order over the neutrino mass parameter $b$
contribution to the magnetic moment, we find out that the sum of
the corresponding contributions of Eqs.~\eqref{mu1}-\eqref{mu56}
to the coefficient $\bar{\mu}_{1}(a,\alpha)$ is given by
\begin{equation}
  \label{mu1tot}
  \bar{\mu}_{1}(a,\alpha)=
  \sum_{i=1}^{6}\bar{\mu}^{(i)}_{1}(a,\alpha)=
  {\frac{1}{12(1-a)^{5}}}
  (5-26a+6a\ln a-36a^{2}-60a^{2}\ln a+58a^{3}-18a^{3}\ln a-a^{4}).
\end{equation}
Thus, we explicitly show by Eqs.~\eqref{mu0tot} and \eqref{mu1tot}
that in the one-loop level and to the second order in the
expansion over the neutrino mass parameter $b$ the neutrino
magnetic moment is a gauge-independent quantity.

The obtained Eqs.~\eqref{mu1}-\eqref{mu56} also enable us to
consider the magnetic moment of a rather heavy neutrino since the
mass parameters $a$ and $b$ are arbitrary in these equations. Let
the neutrino mass  $m_{\nu}$ be much greater than the charged
lepton mass $m_{\ell}$ (this case amounts to $b\gg a$).
Approaching the limit $a\to 0$ in
Eqs.~\eqref{mu0tot}-\eqref{mu1tot}, while keeping $b$ constant,
for the neutrino magnetic moment we receive
\begin{equation}
  \mu={\frac{3eG_{F}}{8\pi^{2}\sqrt{2}}}m_{\nu}
  \left\{
  1+{\frac{5}{18}}b+\dotsb
  \right\}.
\end{equation}

The recent LEP data require that the number of light neutrinos
coupled to $Z$ boson is exactly three. Any additional neutrinos
must be heavier than $80\thinspace\text{GeV}$ (see, e.g.,
Ref.~\cite{Acc99}). Using our formulas
Eqs.~\eqref{mu1}-\eqref{mu56} for a massive Dirac neutrino
magnetic moment we can also examine the case of a very heavy
neutrino. Let us consider the case of neutrino mass being even
greater than $W$ boson mass. To examine this situation we should
fix the gauge parameter $\alpha$ in Eqs.~\eqref{mu1}-\eqref{mu56}
for simplicity of the computations. In what follows we set
$\alpha=1$ that corresponds to the 't~Hooft-Feynman gauge. Thus,
for the sum of all the contributions to the magnetic moment we
obtain the expression
\begin{gather}
  \label{musuperheavy}
  \bar{\mu}=
  \delta_\ell
  \int_0^1 dz
  \frac{1}{\mathcal{D}}-
  (1-2\delta_W+3\delta_\ell)
  \frac{1}{2}
  \int_0^1 dz
  \thinspace z
  \frac{1}{\mathcal{D}}+
  (1+2\delta_W+\delta_\ell)
  \frac{1}{2}
  \int_0^1 dz
  \thinspace z^2
  \frac{1}{\mathcal{D}},
  \\
  \mathcal{D}=z^2-z(1-\delta_W+\delta_\ell)+\delta_\ell,
  \notag
\end{gather}
where we redefined our mass parameters and introduced the two new
quantities: $\delta_W=1/b=(M_W/m_\nu)^2$ and
$\delta_\ell=a/b=(m_\ell/m_\nu)^2$. The case of the super heavy
neutrino corresponds to the values of the new mass parameters in
the range $\delta_\ell\ll\delta_W\ll 1$.

One can prove by means of the direct calculation that
\begin{equation}
  \label{superheavylim}
  \lim_{\delta_W\to 0}
  \delta_W \mathfrak{I}_n =0,
  \quad
  \text{and}
  \quad
  \lim_{\delta_\ell\to 0}
  \delta_\ell \mathfrak{I}_n =0,
  \quad
  n=0,\dots,2,
\end{equation}
where
\begin{equation}
  \label{superheavyI}
  \mathfrak{I}_n=
  \int_0^1 dz
  z^n
  \frac{1}{\mathcal{D}}.
\end{equation}
Using Eqs.~\eqref{superheavylim} and \eqref{superheavyI}, we find
that the function $\bar{\mu}$ in Eq.~\eqref{musuperheavy} is equal
to $1/2$ that corresponds to the magnetic moment
\begin{equation}
  \label{mushfin}
  \mu={\frac{eG_{F}}{8\pi^{2}\sqrt{2}}}m_{\nu}.
\end{equation}
Eq.~\eqref{mushfin} presents the magnetic moment of a heavy
neutrino with the mass much greater than $W$ boson mass.

At the end of this section let us compare the calculation of the
neutrino magnetic moment in the unitary and $R_{\xi}$ gauges. The
calculations performed within these gauges, as it was mentioned in
Ref.~\cite{FujLeeSan72}, are formally equivalent, i.e. the two
Feynman amplitudes become equal if we approach the limit
$\alpha\to\infty$ prior corresponding loop integrals are carried
out. The diagrams involving unphysical scalar bosons must
disappear in the unitary gauge. Therefore, in such diagrams the
limit $\alpha\to\infty$ and the integration over virtual momenta
must be commuting procedures. We directly verified this statement
for particular case of the calculation of the massive neutrino
magnetic moment. Indeed, on the basis of either exact
formulae~\eqref{mu1}-\eqref{mu56} or the expansions given by
Eqs.~\eqref{mu01}-\eqref{mu056}, we find out that the
contributions of the diagrams depicted in Fig.~\ref{prvertb},
\ref{prvertc}, \ref{prverte}, and \ref{prvertf}, which involve the
scalar boson, vanish in the limit $\alpha\to\infty$.

\subsection{\label{fMq2} Neutrino Magnetic Form Factor at
Non-zero Momentum Transfer}

In this subsection we study the neutrino magnetic form factor at
non-zero momentum transfer in arbitrary $R_{\xi}$ gauge as well as
for arbitrary charged lepton and neutrino mass parameters ($a$ and
$b$). The $\gamma-Z$ self-energy diagrams shown in
Fig.~\ref{gZverta}-\ref{gZverth} also do not contribute to the
magnetic form factor at $q^{2}\not=0$. Therefore, the total
one-loop value for the neutrino magnetic form factor is given by
\begin{equation*}
  f_{M}(q^{2})=\frac{eG_{F}}{4\pi^{2}\sqrt{2}}m_{\nu}
  \sum_{i=1}^{6}\bar{f}_{M}^{(i)}(q^{2}),
\end{equation*}
where $\bar{f}_{M}^{(i)}(q^{2})$ are the contributions to the
magnetic moment from the corresponding diagrams shown in
Fig.~\ref{prverta}-\ref{prvertf}. For the coefficients
$\bar{f}_{M}^{(i)}(q^{2})$ we have
\begin{multline}
  \label{fM1}
  \bar{f}_{M}^{(1)}(q^{2})=
  \int_{0}^{1}dz
  \int_{0}^{z}dy(2-3z+z^{2})\frac{1}{D_{1}}-
  \\
  \frac{1}{2}\int_{0}^{1}dz
  \int_{0}^{z}dy(az^{2}-bz^{2}(1-z)-ty(z-y)(2-z))
  \left[\frac{1}{D_{1}(\alpha)}-\frac{1}{D_{1}}\right]+
  \\
  \frac{1}{2}\int_{0}^{1}dz
  \int_{0}^{z}dy(2-3z)
  \left[\ln D_{1}(\alpha)-\ln D_{1}\right],
\end{multline}
\begin{equation}
  \label{fM2}
  \bar{f}_{M}^{(2)}(q^{2})=
  \frac{1}{2}\int_{0}^{1}dz
  \int_{0}^{z}dy\thinspace z(a+az-b(1-z))
  \frac{1}{D_{1}(\alpha)},
\end{equation}
\begin{equation}
  \label{fM3}
  \bar{f}_{M}^{(3)}(q^{2})=
  \frac{1}{2}\int_{0}^{1}dz
  \int_{0}^{z}dy(2a-3az+az^{2}-bz(1-z))
  \frac{1}{D_{2}(\alpha)},
\end{equation}
\begin{multline}
  \label{fM4}
  \bar{f}_{M}^{(4)}(q^{2})=
  \frac{1}{2}\int_{0}^{1}dz
  \int_{0}^{z}dy\thinspace z(1+2z)
  \frac{1}{D_{2}}+
  \\
  \frac{1}{2}\int_{0}^{1}dz
  \int_{0}^{z}dy
  (b(1-z)^{2}(z(1-z)-2y)-
  ty(z-y)(2y-3z+z^{2})-2ty)
  \left[
  \frac{1}{D_{2}(\alpha)+y(1-\alpha)}-\frac{1}{D_{2}}
  \right]+
  \\
  \frac{1}{2}\int_{0}^{1}dz
  \int_{0}^{z}dy (-2+9z-4z^{2}-6y)
  \left[
  \ln \left(D_{2}(\alpha)+y(1-\alpha)\right)-\ln D_{2}
  \right]-
  \\
  \frac{t}{4}\int_{0}^{1}dz
  \int_{0}^{z}dy
  (bz(1-3z+z^{2}+z^{3})-ty(z-y)(2-z-z^{2}))
  \left[
  \frac{1}{D_{2}}+\frac{1}{D_{2}(\alpha)}-
  \frac{2}{D_{2}(\alpha)+y(1-\alpha)}
  \right]+
  \\
  \frac{t}{8}\int_{0}^{1}dz
  \int_{0}^{z}dy (8-13z+3z^{2})
  \left[
  \ln D_{2}+\ln D_{2}(\alpha)-
  2\ln \left(D_{2}(\alpha)+y(1-\alpha)\right)
  \right],
\end{multline}
\begin{multline}
  \label{fM56}
  \bar{f}_{M}^{(5)+(6)}(q^{2})=
  \int_{0}^{1}dz
  \int_{0}^{z}dy
  \thinspace y\frac{1}{D_{2}(\alpha)+y(1-\alpha)}+
  \\
  \frac{1}{2}\int_{0}^{1}dz
  \int_{0}^{z}dy
  ((a-bz)(1-z)^{2}+ty(z-y)(1-z))
  \left[
  \frac{1}{D_{2}(\alpha)+y(1-\alpha)}-\frac{1}{D_{2}(\alpha)}
  \right]+
  \\
  \frac{1}{2}\int_{0}^{1}dz
  \int_{0}^{z}dy (2-3z)
  \left[
  \ln \left(D_{2}(\alpha)+y(1-\alpha)\right)-\ln D_{2}(\alpha)
  \right],
\end{multline}
where
\begin{align*}
  D_{1}(\alpha) & =\alpha+(a-\alpha)z-bz(1-z)+ty(z-y),
  &
  D_{1}=D_{1}(\alpha=1) & =1+(a-1)z-bz(1-z)+ty(z-y),
  \\
  D_{2}(\alpha) & =a+(\alpha-a)z-bz(1-z)+ty(z-y),
  &
  D_{2}=D_{2}(\alpha=1) & =a+(1-a)z-bz(1-z)+ty(z-y),
\end{align*}
and $t=-q^{2}/M_{W}^{2}$.

We discuss below the large positive $t$ behavior of the integrals
in the expressions of the proper vertices contributions to
$f_{M}(q^{2})$. For example, let us consider the following
integral at $t\to+\infty$:
\begin{equation}
  \label{calI}
  J(t)=t\int_{0}^{z}dy\frac{y}{D_{2}(\alpha)}=
  \int_{0}^{z}dy\frac{y}{(y-y_{2})(y_{1}-y)},
\end{equation}
where
\begin{equation}
  \label{y1y2}
  y_{1}=z+\frac{D}{zt}+\dotsb,
  \quad
  y_{2}=-\frac{D_{\alpha}}{zt}+\dotsb.
\end{equation}
Performing the integrations we readily find that
\begin{equation}
  \label{calIlim}
  J(t)\to\ln t-\ln D.
\end{equation}
In Eq.~\eqref{calIlim} we dropped the terms like $1/t$ and $(\ln
t)/t$ which are negligible for large positive $t$. The remaining
integrals are evaluated in a similar way. Finally, we find that
\[
  \bar{f}_{M}(t)=\sum_{i=1}^{6}\bar{f}^{(i)}_{M}(t)\to 0,
  \thickspace
  \text{if}
  \thickspace
  t\to+\infty.
\]
The behavior of the magnetic form factor at large negative
$q^{2}$, described above, is consistent with the general Weinberg
theorem \cite{Wei96p505}. However, the case of the massive
neutrino magnetic form factor has never been discussed previously.

It should be noted that in derivation of
Eqs.~\eqref{calI}-\eqref{calIlim} we assumed that $\alpha<\infty$.
Therefore, our result that $\bar{f}_{M}(t)\to 0$ at $t\to+\infty$
is valid in any gauge except the unitary one. The value of
$\bar{f}_{M}(t\to+\infty)$ may not be equal to zero if we at first
set $\alpha=\infty$ and then approach the limit $t\to+\infty$. The
analysis of the large negative $q^{2}$ behavior of magnetic form
factor within the Weinberg-Salam model in the unitary gauge is
given, for instance, in Ref.~\cite{FujLeeSan72}.

\begin{figure}
  \includegraphics[scale=.8]{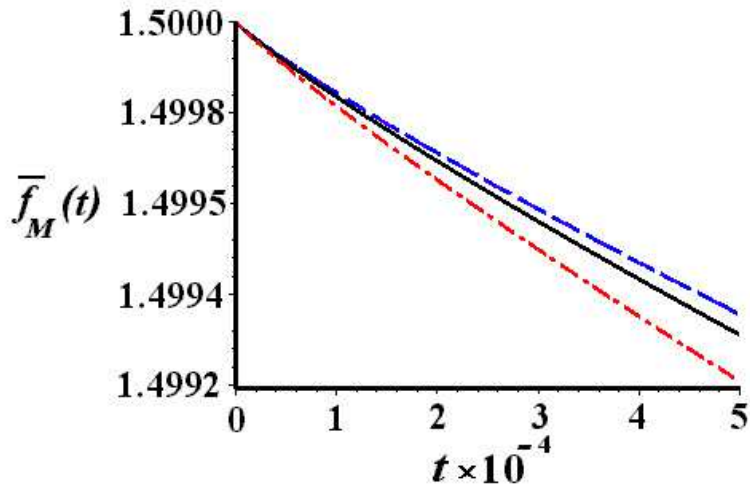}
    \caption{The massive neutrino magnetic form factor
    versus squared transverse momentum in different gauges. The dashed
    line corresponds to $\alpha=100$, the solid line to the
    t'Hooft-Feynman gauge ($\alpha=1$), and the dash-dotted line to
    $\alpha=0.1$}
    \label{fMdifg}
\end{figure}

Using the explicit formulae for the massive neutrino magnetic form
factor for arbitrary gauge parameter $\alpha$
[Eqs.~\eqref{fM1}-\eqref{fM56}] we present in Fig.~\ref{fMdifg}
the behavior of the function $\bar{f}_{M}(t)$ for different gauges
and a wide range of $t$: $0\leq t\leq 5\times 10^{-4}$.
It can be seen that the magnetic form factor becomes gauge
independent at $t=0$ that amounts to the case of on-shell photon.
The value $f_{M}(t=0)$ is equal to the neutrino magnetic moment,
and Fig.~\ref{fMdifg} shows gauge independence of this quantity in
agreement with our exact calculations performed above.

\section{Conclusion \label{concl}}

We have considered the massive neutrino electric charge and
magnetic moment within the context of the standard model supplied
with $\mathrm{SU}(2)$-singlet right-handed neutrino in general
$R_{\xi}$ gauge. Using the dimensional-regularization scheme, we
have calculated the one-loop contributions to the neutrino
electromagnetic vertex function taking exactly into account the
neutrino mass. We have presented the results of our calculations
of different contributions to the neutrino electric charge and
magnetic moment as the closed integral expressions. It has allowed
us to determine the dependencies of these contributions on the
neutrino and corresponding charged lepton masses as well as on the
gauge-fixing parameters. The integral expressions for the neutrino
electric charge and magnetic moment obtained in this work contain
at most two definite integrals which, in principle, can be
performed and expressed in terms of elementary functions. However,
the results are quite cumbersome and therefore we have presented
them as the expansion over the neutrino mass parameter $b$. For
several diagrams, which contribute to the neutrino charge and
magnetic moment and which have been calculated in
Ref.~\cite{CabBerVidZep00} with mistakes, we have found the
correct results.

We have found the general expressions for the contributions to
neutrino electric form factor. These formulae have been derived in
general $R_\xi$ gauge and at arbitrary value of $q^2$. We have
shown that the electric charge of a massive neutrino is a gauge
independent and vanishing parameter in the first two orders of the
expansion over the neutrino mass parameter $b$. In the particular
choice of the 't~Hooft-Feynman gauge we have also demonstrated
that the neutrino charge is zero in all orders of expansion over
$b$, i.e. for arbitrary mass of neutrino. In the previously
published works devoted to the calculation of the neutrino
electric charge the case of massless neutrino was studied within
the Georgi-Glashow (see Ref.~\cite{FujLeeSan72}) and
Weinberg-Salam (see
Refs.~\cite{BarGasLau72,MarSir80,Sak81,LucRosZep84}) models.
However, it is clear that the massless particle must be
electrically neutral. Although there is no doubt that the massive
neutrino electric charge must be also zero, however it has not
been yet shown how it actually happens for corresponding Feynman
diagrams.

There are other reasons to prove by the direct calculations that
the value of the massive neutrino electric charge is zero. For
example, this problem is important in consideration of the
neutrino spin oscillations. In the series of our works
\cite{EgoLobStu00,LobStu01,DvoStu02JHEP,LobStu03} we have
elaborated the quasi-classical approach for the description of the
neutrino spin oscillations in arbitrary external electromagnetic
field. An essential point in that studies has been the zero charge
of the massive neutrino. In this paper we have substantiated this
assumption.

From the obtained closed two-integral expression for the massive
neutrino electric form factor it is also possible to derive the
neutrino charge radius.

The structure of the massive neutrino electromagnetic vertex
function have been examined in this work. We have directly
verified the decomposition of the neutrino vertex function. It has
been found out that the value of the additional "form factor"
$f_{5}(q^{2})$, which is proportional to $\gamma_{\mu}\gamma_{5}$
matrix, at $q^{2}=0$ is zero in the first two orders of the
expansion over the neutrino mass parameter $b$ and for arbitrary
gauge parameter $\alpha$. The vanishing of the additional "form
factor" $f_{5}(q^{2})$ in the particular gauge
($\alpha_{Z}=\infty$ and $\alpha_{W}=1$) but for the arbitrary
charged lepton $a$ and neutrino $b$ mass parameters as well as for
arbitrary $q^{2}$ has also been demonstrated. Such direct
calculations have never been carried out before.

For each of the diagrams contributing to the neutrino magnetic
moment, we have obtained the expressions accounting for the
leading (zeroth) and next-to-leading (first) orders in the
neutrino mass parameter $b$ with the gauge dependence shown
explicitly. Each of the contributions is finite and the sum of all
contributions turns out to be gauge-independent. Our calculations
also enable us to get the neutrino magnetic moment in the
following ranges of the neutrino, charged lepton, and $W$ boson
masses: $m_\nu\ll m_\ell\ll M_W$, $m_\ell\ll m_\nu\ll M_W$, and
$m_\ell\ll M_W\ll  m_\nu$, which span almost all the cases
presently discussed within different theoretical models. We have
also presented the general formulae for the massive neutrino
magnetic form factor at arbitrary $q^{2}$.

As for the behavior of the neutrino magnetic form factor at
$q^{2}{\not=}0$, we have found that the function $f_{M}(q^{2})$
essentially depends on the gauge fixing parameter $\alpha$ at
$q^{2}{\not=}0$. The magnetic form factor may depend on the gauge
parameter at $q^{2}{\not=}0$ since it is not a measurable property
of a particle and, therefore, may not be invariant under the gauge
group transformations. The consideration of the gauge parameter
dependence of the neutrino magnetic form factor as well as its
asymptotic behavior at large negative $q^{2}$ in the limit $b\to
0$ within the Weinberg-Salam model is presented in
Ref.~\cite{FujLeeSan72}. The transition magnetic moment of the
Dirac neutrinos coupled with the light fermion $f$ ($m_{f}\approx
m_{\nu}$) and with the light scalar boson $\phi$ ($m_{\phi}\ll
m_{\nu}-m_{f}$) through the Yukawa interaction $\bar{\nu}f\phi$ is
discussed in Ref.~\cite{FreNevVys97}. The transition magnetic
moment dependence on $q^{2}$ is also considered there. The
analysis of the neutrino magnetic moment is presented in
Ref.~\cite{CzaGluZra99} for various versions of the left-right
symmetric models. The results of our massive neutrino magnetic
moment calculations can be applied to the treatment of the
magnetic moment (including the transitional magnetic moment)
within the left-right symmetric model.

Although we have not studied the neutrino anapole form factor in
this paper, this particular problem (which we discuss in
\cite{DvoStuUP}) is also important since, for instance, the
anapole moment (the value of the anapole form factor at $q^{2}=0$)
is the only static electromagnetic property of a Majorana neutrino
(see, for instance, Refs.~\cite{Kay82,DubKuz98}). It should be
noted that even a massless particle can posses the anapole moment,
unlike the magnetic moment. Some resent papers are worth
mentioning in this respect (see \cite{BukDubKuz98,Ros00} and
references therein). However, the investigation of the neutrino
anapole moment faces serious difficulties such as its
observability and gauge dependence.

\begin{acknowledgments}
We should like to thank A.~Lobanov and A.~Pivovarov for useful
discussions. We are also thankful J.~Gluza for comments on our
paper and proposals for the further research that can be done in
application of our studies to the case of the left-right symmetric
models. The authors are grateful to J.~Bernab\'{e}u, L.~Cabral and
J.~Vidal for the discussion on the discrepancy between our and
their results after which the results have coincided. We are also
indebted to K.~Stepaniants for helpful comments on the analytical
calculations.
\end{acknowledgments}

\appendix

\section{Feynman Rules\label{FeynRul}}

In this Appendix we present the full list of the Feynman rules
\cite{AokHioKawKonMut82} necessary for the calculation of the
massive neutrino electromagnetic vertex. In the $R_{\xi}$ gauge
the propagators for the vector bosons, $W$ and $Z$, an unphysical
charged scalar boson, $\chi$, as well as charged ghosts, $c$ and
$\bar c$, are presented in the following form
\begin{align*}
  D_{\mu\nu}^{(W)}(k) &= \frac{1}{k^{2}-M_{W}^{2}+i\epsilon}
  \left[
  g_{\mu\nu}-(1-\alpha) \frac{k_{\mu}k_{\nu}}
  {k^{2}-\alpha M_{W}^{2}+i\epsilon}
  \right],
  \\
  D_{\mu\nu}^{(Z)}(k) &= \frac{1}{k^{2}-M_{Z}^{2}+i\epsilon}
  \left[
  g_{\mu\nu}-(1-\alpha_{Z})
  \frac{k_{\mu}k_{\nu}}{k^{2}-\alpha_{Z}M_{Z}^{2}+i\epsilon}
  \right],
\end{align*}
\begin{equation*}
  D^{(\chi)}(k) = \frac{1}{\alpha M_{W}^{2}-k^{2}-i\epsilon},
  \quad
  D^{(c)}(k)=D^{(\overline{c})}(k)=
  \frac{1}{\alpha M_{W}^{2}-k^{2}-i\epsilon}.
\end{equation*}
The fermion propagator has the standard form
\begin{equation*}
  S(k)=\frac{{\not k}+m_{n}}{m_{n}^{2}-k^{2}-i\epsilon},
\end{equation*}
where $n$ denotes the type of a fermion.

All vertices can be divided into several classes. We append below
the corresponding graphs and Feynman rules for each of these
classes.
\begin{figure}[H]
\begin{center}
  \subfigure[\hfill$g\cos\theta_{W}
       \{(k-p)^{\gamma}g^{\alpha\beta}+$
       $(p-q)^{\alpha}g^{\beta\gamma}+
       (q-k)^{\beta}g^{\gamma\alpha}\}$]
  {\label{3vba}
  \includegraphics{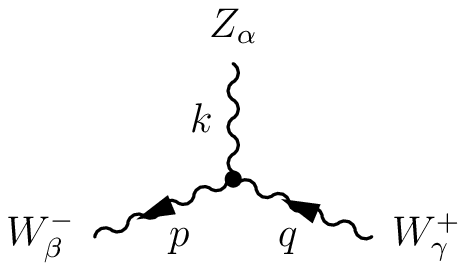}}
    \hspace{2cm}
  \subfigure[\hfill$e\{(k-p)^{\gamma}g^{\alpha\beta}+$
       $(p-q)^{\alpha}g^{\beta\gamma}+
       (q-k)^{\beta}g^{\gamma\alpha}\}$]
  {\label{3vbb}
  \includegraphics{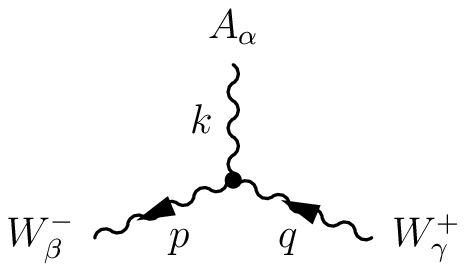}}
    \caption{\subref{3vba}-\subref{3vbb}
    vector boson triplex vertices.}
\end{center}
\end{figure}
\begin{figure}[H]
\begin{center}
  \subfigure[\hfill$eg\cos\theta_{W}\times$
    $\{g^{\alpha\gamma}g^{\beta\delta}+
    g^{\alpha\delta}g^{\beta\gamma}-
    2g^{\alpha\beta}g^{\gamma\delta}\}$]
  {\label{4vb}
  \includegraphics{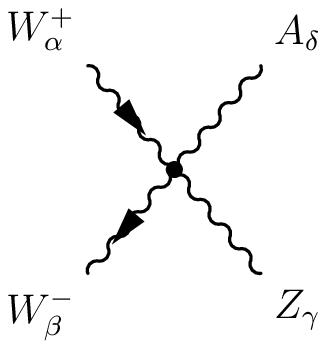}}
    \caption{\subref{4vb} vector boson quadruple vertex.}
\end{center}
\end{figure}
\begin{figure}[H]
\begin{center}
  \subfigure[$(g/\sqrt{2})\gamma_{\alpha}^{L}$]
  {\label{1vb2fa}
  \includegraphics{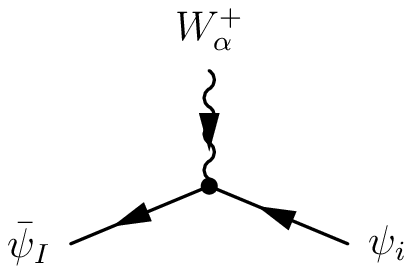}}
    \hspace{2cm}
  \subfigure[$(g/\sqrt{2})\gamma_{\alpha}^{L}$]
  {\label{1vb2fb}
  \includegraphics{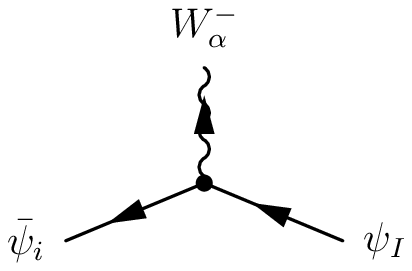}}
    \\
  \subfigure[$eQ_{n}\gamma_{\alpha}^{L}$, $n=i$, $I$]
  {\label{1vb2fc}
  \includegraphics{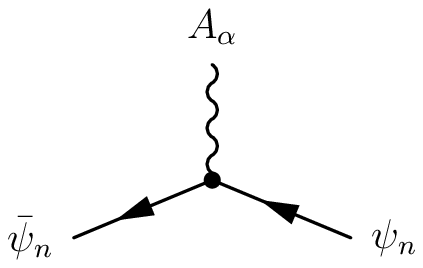}}
    \\
  \subfigure[\hfill$(g/2\cos\theta_{W})\gamma_{\alpha}\times$
      $\big(1/2-2Q_{I}\sin^{2}\theta_{W}+\gamma_{5}/2
      \big)$]
  {\label{1vb2fd}
  \includegraphics{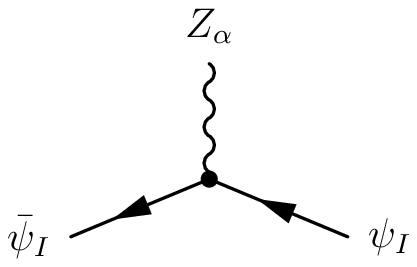}}
    \hspace{2cm}
  \subfigure[\hfill$-(g/2\cos\theta_{W})\gamma_{\alpha}\times$
      $\big(1/2+2Q_{i}\sin^{2}\theta_{W}+\gamma_{5}/2
      \big)$]
  {\label{1vb2fe}
  \includegraphics{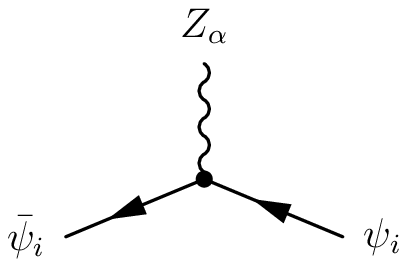}}
    \caption{\subref{1vb2fa}-\subref{1vb2fe}
    one vector boson and two fermions vertices.}
\end{center}
\end{figure}
\begin{figure}[H]
\begin{center}
  \subfigure[\hfill$-(g/2\cos\theta_{W})\times$
    $\cos2\theta_{W}(p-q)_{\alpha}$]
  {\label{1vb2sba}
  \includegraphics{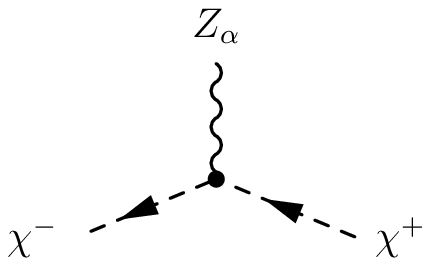}}
    \hspace{2cm}
  \subfigure[$-e(p-q)_{\alpha}$]
  {\label{1vb2sbb}
  \includegraphics{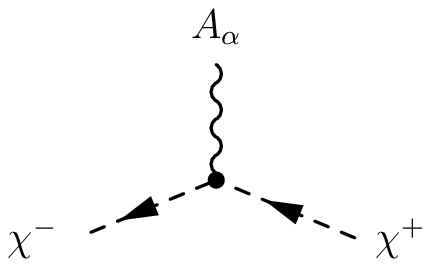}}
    \caption{\subref{1vb2sba}-\subref{1vb2sbb}
    one vector boson and two scalar bosons vertices.}
\end{center}
\end{figure}
\begin{figure}[H]
\begin{center}
  \subfigure[$ig\sin^{2}\theta_{W}M_{Z}g_{\alpha\beta}$]
  {\label{2vb1sba}
  \includegraphics{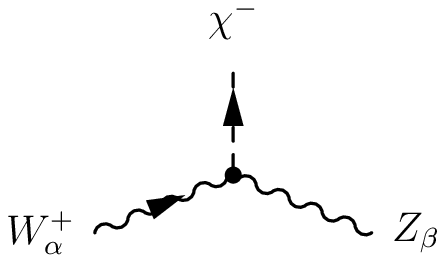}}
    \hspace{2cm}
  \subfigure[$-ieM_{W}g_{\alpha\beta}$]
  {\label{2vb1sbb}
  \includegraphics{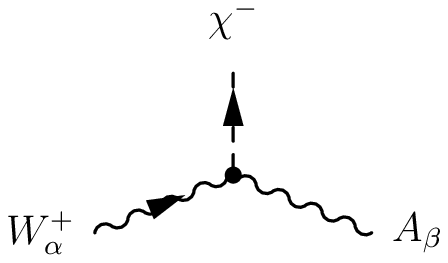}}
    \\
  \subfigure[$-ig\sin^{2}\theta_{W}M_{Z}g_{\alpha\beta}$]
  {\label{2vb1sbc}
  \includegraphics{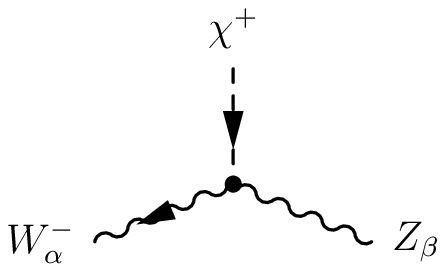}}
    \hspace{2cm}
  \subfigure[$ieM_{W}g_{\alpha\beta}$]
  {\label{2vb1sbd}
  \includegraphics{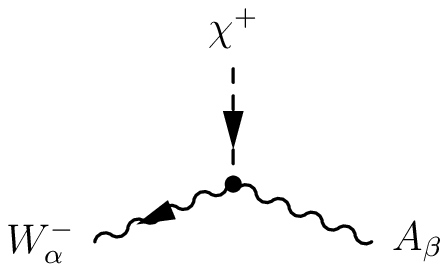}}
    \caption{\subref{2vb1sba}-\subref{2vb1sbd}
    two vector bosons and one scalar boson vertices.}
\end{center}
\end{figure}
\begin{figure}[H]
\begin{center}
  \subfigure[\hfill$-i(g/\sqrt{2}M_{W})\times$
  $(m_{i}P_{R}-m_{I}P_{L})$]
  {\label{1sb2fa}
  \includegraphics{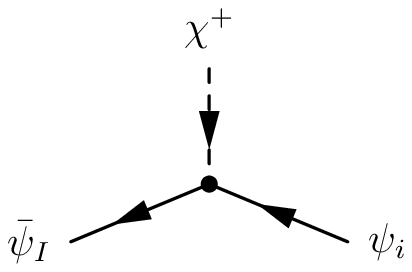}}
    \hspace{2cm}
  \subfigure[\hfill$-i(g/\sqrt{2}M_{W})\times$
  $(m_{I}P_{R}-m_{i}P_{L})$]
  {\label{1sb2fb}
  \includegraphics{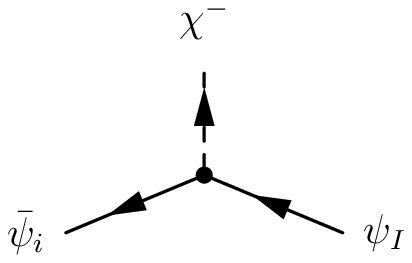}}
    \caption{\subref{1sb2fa}-\subref{1sb2fb}
    one scalar boson and two fermions vertices.}
\end{center}
\end{figure}
\begin{figure}[H]
\begin{center}
  \subfigure[$-g\cos \theta_{W}p_{\alpha}$]
  {\label{1vb2cga}
  \includegraphics{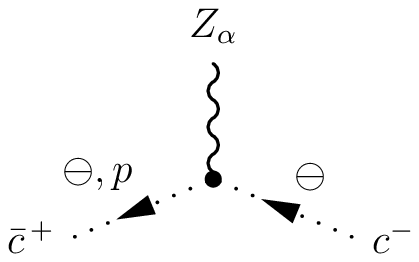}}
    \hspace{2cm}
  \subfigure[$g\cos \theta_{W}p_{\alpha}$]
  {\label{1vb2cgb}
  \includegraphics{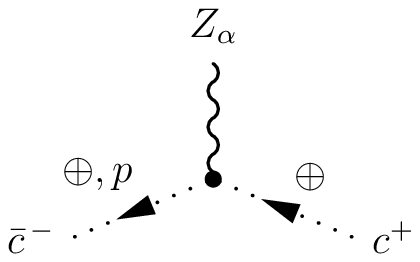}}
    \\
  \subfigure[$-ep_{\alpha}$]
  {\label{1vb2cgc}
  \includegraphics{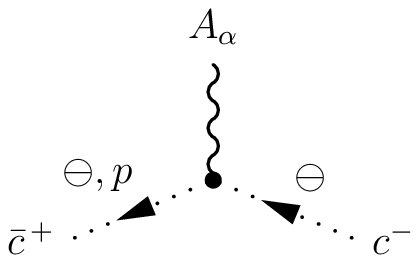}}
    \hspace{2cm}
  \subfigure[$ep_{\alpha}$]
  {\label{1vb2cgd}
  \includegraphics{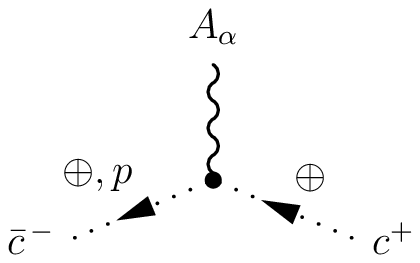}}
    \caption{\subref{1vb2cga}-\subref{1vb2cgd}
    one vector boson and two charged ghosts vertices.}
\end{center}
\end{figure}
\begin{figure}[H]
\begin{center}
  \subfigure[\hfill$eg\thinspace g_{\alpha\beta}\times$
  $\cos2\theta_{W}/\cos\theta_{W}$]
  {\label{2vb2sba}
  \includegraphics{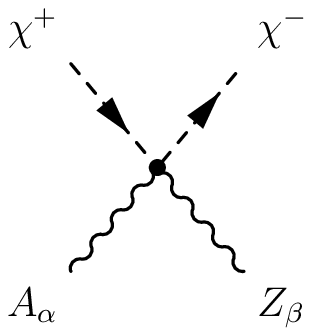}}
    \caption{\subref{2vb2sba} two vector bosons and two scalar bosons vertex.}
\end{center}
\end{figure}
All the momenta of particles associated with vertices are taken to
flow in. $Q_{i,I}$ represent electromagnetic charges of the fields
$\psi_{i,I}$ in the units of $e$. $\psi_{i,I}$ representing the
three generations of leptons and quarks correspond to usual "up"
(all types of neutrinos as well as $u$, $c$ and $t$ quarks;
$I_{3}=+1/2$) and "down" (all types of leptons as well as $d$, $s$
and $b$ quarks; $I_{3}=-1/2$) components of an isodoublet,
respectively, $I_{3}$ is the third component of the isospin.

The arrow on a line indicates the direction of the flow of a
certain quantum number: the charge for $W^{\pm}$, $\chi^{\pm}$,
the fermion number for $\psi$, the ghost number for $c$,
$\overline{c}$. The symbol $\oplus$ or $\ominus$ at the charged
ghost lines stands for the sign of the charge carried by the
arrow.

\section{Feynman integrals\label{FeynInt}}

In our calculation of Feynman integrals over virtual momenta we
use dimensional-regularization scheme with the following natural
properties of $\gamma$-matrix algebra:
\begin{equation*}
  \{\gamma_{\mu},\gamma_{5}\}=0,\quad
  \{\gamma_{\mu},\gamma_{\nu}\}=2g_{\mu\nu},\quad
  g^{\mu\nu}g_{\mu\nu}=N,
\end{equation*}
where $N=4-2\varepsilon$ is the number of dimensions.

The dimensional regularization of the loop integrals in the
Euclidian space is performed in the following way:
\begin{equation*}
  \frac{1}{(2\pi)^{4}}\int d^{4}k\to \frac{1}{(2\pi)^{N}}\int d^{N}k
  \equiv \frac{\lambda^{2\varepsilon}}{(2\pi)^{N}}
  \int\limits_{\Omega(N)} d\bm{\Omega}
  \int\limits_{0}^{\infty} k^{N-1}dk,
\end{equation*}
where $\Omega(N)=2\pi^{N/2}/\Gamma(N/2)$ is the area of a unit
sphere in $N$ dimensions. The dependence of an arbitrary positive
parameter $\lambda$, which has the mass dimensionality, is
introduced to provide the total dimensionality of an integral. The
general technique for calculation of various loop integrals in the
dimensional regularization scheme can be found, for example, in
Ref.~\cite{BogShi80}. It should be, however, rather helpful to
include here some of the typical loop integrals which one
encounters while calculating the electromagnetic vertex function,
\begin{gather*}
  F_{L}^{(0)}={\frac{i}{\pi^{2}}}
  \int d^{N}k {\frac{1}{(k^{2}+X)^{L}}}=
  -\left(
  {\frac{\lambda^{2}i^{2}}{\pi}}
  \right)
  ^{\varepsilon}
  {\frac{\Gamma(L-2+\varepsilon)}{\Gamma(L)}}
  {\frac{1}{X^{L-2+\varepsilon}}},
  \\
  F_{1}^{(0)}=X
  \left[
  {\frac{1}{\varepsilon}}-\ln
  \left(
  -{\frac{\pi X}{\lambda^{2}}}
  \right)-{\mathbb C}+1
  \right],
  \quad
  F_{2}^{(0)}=
  -{\frac{1}{\varepsilon}}+\ln
  \left(
  -{\frac{\pi X}{\lambda^{2}}}
  \right)+{\mathbb C},
  \\
  F_{3}^{(0)}=-{\frac{1}{2X}},
  \quad
  F_{4}^{(0)}=-{\frac{1}{6X^{2}}},
  \quad
  F_{5}^{(0)}=-{\frac{1}{12X^{3}}},
\end{gather*}
\begin{gather*}
  F_{L}^{(1)}={\frac{i}{\pi^{2}}}
  \int d^{N}k {\frac{k^{2}}{(k^{2}+X)^{L}}}=
  \left(
  {\frac{\lambda^{2}i^{2}}{\pi}}
  \right)
  ^{\varepsilon}
  {\frac{\Gamma(L-3+\varepsilon)}{\Gamma(L)}}
  {\frac{\varepsilon-2}{X^{L-3+\varepsilon}}},
  \\
  F_{2}^{(1)}=2X
  \left[
  {\frac{1}{\varepsilon}}-\ln
  \left(
  -{\frac{\pi X}{\lambda^{2}}}
  \right)-{\mathbb C}+{\frac{1}{2}}
  \right],
  \quad
  F_{3}^{(1)}=
  -{\frac{1}{\varepsilon}}+\ln
  \left(
  -{\frac{\pi X}{\lambda^{2}}}
  \right)+{\mathbb C}+{\frac{1}{2}},
  \\
  F_{4}^{(1)}=-{\frac{1}{3X}},
  \quad
  F_{5}^{(1)}=-{\frac{1}{12X^{2}}},
\end{gather*}
\begin{gather*}
  F_{L}^{(2)}={\frac{i}{\pi^{2}}}
  \int d^{N}k {\frac{(k^{2})^{2}}{(k^{2}+X)^{L}}}=
  -\left(
  {\frac{\lambda^{2}i^{2}}{\pi}}
  \right)
  ^{\varepsilon}
  {\frac{\Gamma(L-4+\varepsilon)}{\Gamma(L)}}
  {\frac{(\varepsilon-2)(\varepsilon-3)}
  {X^{L-4+\varepsilon}}},
  \\
  F_{3}^{(2)}=3X
  \left[
  {\frac{1}{\varepsilon}}-\ln
  \left(
  -{\frac{\pi X}{\lambda^{2}}}
  \right)-{\mathbb C}+{\frac{1}{6}}
  \right],
  \quad
  F_{4}^{(2)}=
  -{\frac{1}{\varepsilon}}+\ln
  \left(
  -{\frac{\pi X}{\lambda^{2}}}
  \right)+{\mathbb C}+{\frac{5}{6}},
  \\
  F_{5}^{(2)}=-{\frac{1}{4X}},
\end{gather*}
where $\mathbb{C}\approx 0.5772157$ is the Euler constant.

\bibliography{general}

\end{document}